\documentstyle[epsf,epsfig]{mn}

\def\phs{\phantom{$-$}} 
\def\phm#1{\phantom{#1}} 
  
\def\lesssim{\mathrel{\hbox{\rlap{\hbox{\lower4pt\hbox{$\sim$}}}\hbox{$<$}}}} 
\def\gtrsim{\mathrel{\hbox{\rlap{\hbox{\lower4pt\hbox{$\sim$}}}\hbox{$>$}}}}

\newcommand{\mm}[1]{\mbox{$#1$}} 
\newcommand{\unit}[1]{\ifmmode \:\mbox{\rm #1}\else \mbox{#1}\fi} 
 
\renewcommand{\sb}[1]{_{\rm #1}} 
 
\newcommand{\mone}{\mm{^{-1}}}

 \newcommand{\kms}{\unit{km~s\mone}} 
\newcommand{\kpc}{\unit{kpc}} \newcommand{\mpc}{\unit{Mpc}} 
\newcommand{\hkpc}{\mm{h\mone}\kpc} 
\newcommand{\hmpc}{\mm{h\mone}\mpc}

\newcommand{\dn}{\mm{D\sb{n}}} 
\newcommand{\dnsig}{\mm{D\sb{n}-\sigma}} \newcommand{\mg}{Mg$_{2}$}

\newcommand{\eqref}[1]{equation~(\ref{eq:#1})}

\title 
[Galaxy clusters in the Perseus--Pisces region -- I. 
Spectroscopic and photometric data] 
{Galaxy clusters in the Perseus--Pisces region -- I.\\ 
Spectroscopic and photometric data for early-type galaxies} 

\author [R.J. Smith et al.] {R.J. Smith$^1$, J.R. Lucey$^1$, M.J. 
Hudson$^{1,2}\thanks{CITA National Fellow.}$, J. Steel$^1$\\ $^1$ 
Department of Physics, University of Durham, Science Laboratories, 
South Road, Durham DH1 3LE, United Kingdom.\\ $^2$ Department of 
Physics \& Astronomy, University of Victoria, P.O. Box 3055, 
Victoria BC V8W 3PN, Canada (Present address). } 

\begin{document} 
\maketitle 

\begin{abstract} 

We present new spectroscopic and photometric data for 137 early-type 
galaxies in nine clusters, and for a set of nearby standard galaxies. The 
clusters studied are Perseus (A0426), Pisces, A0262, A0347, J8, 
HMS0122+3305, 7S21, A2199 and A2634. 

Our spectroscopic data comprise radial velocities ($cz$), central 
velocity dispersions ($\sigma$) and magnesium line strength indices 
(\mg). Internal errors (derived from repeat observations) are 7.6 
per cent on each measurement of velocity dispersion, and 0.010 mag. 
on each \mg\ measurement. 

Following J\o rgensen et al., we correct our $\sigma$ and 
\mg\ results to a physical aperture size of 1.19\hkpc. We correct 
the major published datasets to the same aperture size, and define a 
new `standard system' by the aperture-corrected Lick data of 
Davies et al. Through extensive intercomparisons with data 
from the literature, we present the corrections required to bring 
the major published datasets onto the standard system. The 
uncertainty in these corrections is computed. We demonstrate that 
our new velocity dispersion data can be brought into consistency 
with the standard system, to an uncertainty of $\lesssim$ 0.01 dex. 

From R-band CCD photometry, we derive effective diameter ($A_{\rm 
e}$), mean surface brightness within effective diameter 
($\langle\mu\rangle_{\rm e}$) and an R-band diameter equivalent to 
the \dn\ parameter of Dressler et al. Internal comparisons 
indicate an average error of 0.005 in each measurement of log\dn. 
The combination $\log A_{\rm e} - 0.3 \langle\mu\rangle_{\rm e}$, 
approximately the quantity used in the Fundamental Plane distance 
indicator, has an 
uncertainty of 0.006 per measurement. The photometric data can be 
brought onto a system consistent with external data at the level of 
0.5 per cent in distance. 

These data will be used in a companion paper, to derive distance and 
peculiar velocity estimates for the nine clusters studied. 

\end{abstract} 

\begin{keywords} 
galaxies: clusters: general --- 
galaxies: elliptical and lenticular, cD --- 
galaxies: distances and redshifts --- 
galaxies: fundamental parameters 
\end{keywords} 

\section{Introduction} \label{intro} 

Streaming motions of galaxies are the only probe of the large-scale 
distribution of mass in the nearby Universe. The dominant large-scale 
concentrations of galaxies within a distance of 8000 \kms\ are the 
Hydra-Centaurus/Great Attractor (hereafter GA) region and the 
Perseus--Pisces (hereafter PP) region (Saunders et al. 1991; Hudson 
1993). 

Strong infall into a massive concentration behind the Cen30 cluster 
was first claimed by Lynden-Bell et al. (1988). While there is 
clearly a coherent streaming motion of galaxies in the direction of 
Centaurus, it remains unclear whether this motion is generated locally 
by the GA, or whether more distant sources are 
responsible. The bulk streaming motion of the PP supercluster allows 
a test of these competing flow models. The GA infall model predicts 
the peculiar velocity of PP to be $\sim -100$ \kms. Alternatively, 
if more distant sources are responsible for the large peculiar motions 
in the Hydra--Centaurus direction, then PP might be expected to take 
part in a similarly large, but negative, bulk motion of $\sim$500\kms.

Previous work on motions in PP has been based mainly on application of 
the Tully \& Fisher (1977) relation to samples of spiral galaxies. 
Using a field-spiral sample, Willick (1990, 1991) claimed that the PP 
supercluster was moving towards the local group (and therefore towards 
the GA) at $441$ \kms. Willick quotes only a random error of 49 \kms\, 
but the study is also subject to a systematic calibration error of 
$\sim$100 \kms. Han \& Mould (1992) analysed a sample of of spirals in 
clusters, and reported an average peculiar motion of $-400 \kms$ for 
PP, in close agreement with Willick. 

As compared to the spiral data, the PP region was not well-sampled in the 
elliptical galaxy survey of Faber et al. (1989, 7S). 
To date, no extensive application of the \dnsig\ / Fundamental Plane 
(FP) method has been conducted in this region. 

In this paper we present new spectroscopic and photometric parameters 
for a sample of early-type galaxies in 7 PP clusters. In a companion 
paper (Hudson et al. 1997; 
hereafter Paper II) we apply the \dnsig\ and 
FP relations to deduce distances and peculiar velocities 
of the clusters. 

Our strategy of observing cluster galaxies is motivated by the 
recognition that a field sample suffers from severe homogeneous and 
inhomogeneous Malmquist bias, particularly in the vicinity of large 
structures such as PP (Hudson 1994). The magnitude of this bias can 
be reduced by grouping galaxies into clusters. The dominance of 
early-type galaxies in cluster cores ensures that samples are fairly 
robust against contamination from the field. 

The acquisition of elliptical galaxy data in the PP region will also 
extend the volume over which one may assess 
the consistency of elliptical galaxy FP/\dnsig\ distances, as 
compared to Tully-Fisher distances for spirals. 
This comparison may reveal that the distance indicator relations are 
affected by systematic variations associated with environmental 
effects or star-formation history (see, for example, Guzm\'an et al. 
1992, Gregg 1995). Kolatt \& Dekel (1994), using a preliminary 
version of the Mark III compilation of velocity data (Willick et al., 
1997), have shown that the motions are consistent with the hypothesis 
that spirals and ellipticals trace the same velocity field. This 
compilation is limited, however, by the less extensive data available 
for ellipticals. The aim of the present work is to provide new, 
high-quality data for ellipticals in PP clusters, for use in mapping 
the velocity field with the FP method. 

The present paper is organised as follows. Section~\ref{selection} 
describes the sample selection. In Section~\ref{spectroscopy}, details 
are given of the spectroscopic observations and data reduction. 
Particular attention is paid to the construction of a `standard 
system' of velocity dispersion measurements, and the estimation of 
systematic errors in the merged data. The photometric data and 
reduction are described in Section~\ref{photometry}. 
Section~\ref{conclusion} concludes the paper with a summary of the 
data quality in terms of random and systematic errors. 

\section{Sample selection} \label{selection} 

\subsection{Selection of cluster sample} 

We define as PP the region of the sky bounded by the limits 
$0^{\rm{h}} < \alpha < 4^{\rm{h}}$ and $+20^{\circ} < \delta < 
+45^{\circ}$. 
It should be noted that this definition is not identical to that of 
Willick (1990, 1991), whose PP region extends from 
$22^{\rm{h}} \lesssim \alpha \lesssim 3^{\rm{h}}$. 
Within this region, the prominent clusters chosen for 
study were: Perseus (A0426), Pisces, A0262, A0347, J8, HMS0122+3305, 
7S21. Of these, J8 (Jackson 1982) lies in the background of the PP ridge, at 
$\sim$10000 \kms, while the remaining six form part of the main body 
of the supercluster, at 4000--6000 \kms. In addition, the clusters 
A2199 and A2634, which do not lie inside the PP region, were observed 
as part of an effort to resolve the 
conflict 
between estimates of their distances (Lucey et al.\ 1991a, 1993, 
1997). Clusters A0262, A2199, A2634 and J8 have also been observed as 
part of the EFAR survey (Wegner et al. 1996).

\begin{figure*} 
\epsfig{file=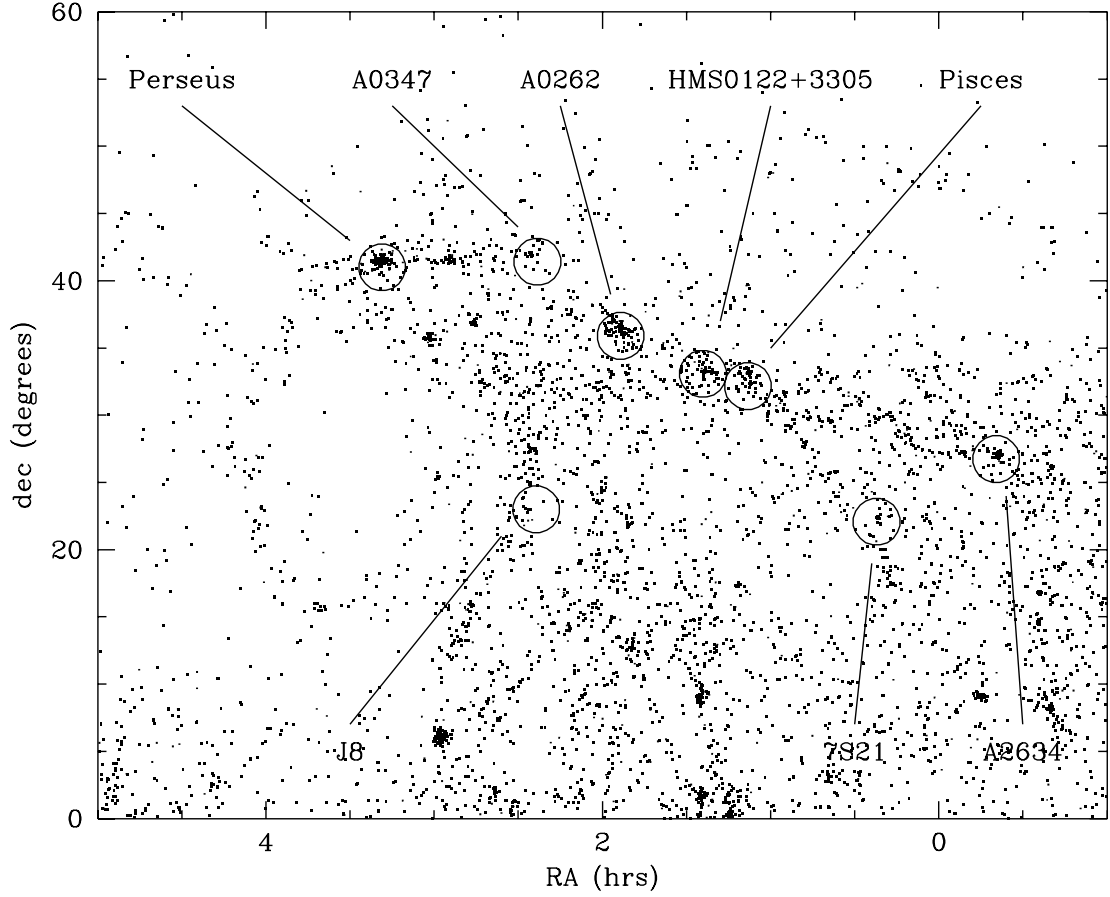,height=160mm,width=160mm} 
\caption{ 
Projected distribution of CfA survey galaxies (with $cz < 12000\ 
\kms$, in the direction of PP. Clusters studied in this work are 
identified by open circles. The circle size is not significant. 
A2199 lies at $\alpha = 16^{\rm{h}}27^{\rm{m}} ; \delta = 
+40^{\circ}$, and is not shown. The low density of galaxies north 
of $+40^{\circ}$ is a result of the limited range of the Arecibo 
radio telescope. East of Perseus, obscuration by the galactic plane 
is apparent.} \label{pponsky} 
\end{figure*}

\begin{figure*} 
\epsfig{file=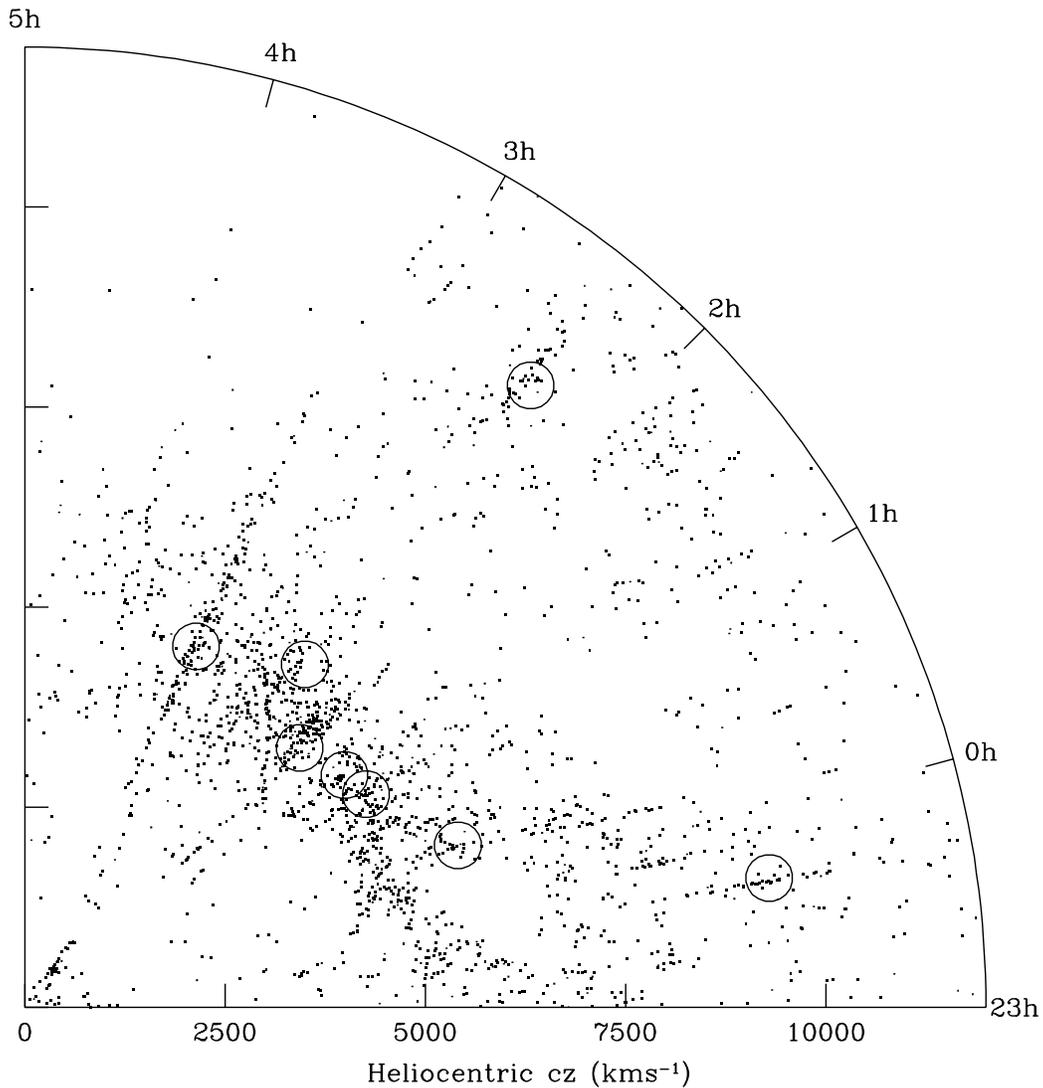,height=160mm,width=160mm} 
\caption{Redshift space distribution of CfA survey galaxies in declination 
range $20^{\circ} < \delta < +45^{\circ}$. Clusters to be studied 
here are marked by open circles. A2199 lies well beyond the limits 
of this plot, at $16.5^{\rm{h}}$ RA.} \label{czspace} 
\end{figure*}

Figure~\ref{pponsky} shows the projected distribution of galaxies in 
the PP region, and slightly beyond in order to show also the position 
of A2634. Galaxy positions are from the CfA redshift survey (Huchra, 
1993). Only those with radial velocities less than 12000 \kms\ are 
plotted. The positions of our target clusters are marked by open 
circles. The redshift-space distribution, for galaxies in 
$+20^{\circ} < \delta < +45^{\circ}$, is illustrated by 
Figure~\ref{czspace}. 

\subsection{Selection criteria for cluster members} 

\begin{table*} 
\caption{Selection criteria for galaxies in each of the PP region clusters. 
$cz_{\rm nom}$ is the CMB-frame redshift used in calculating the 
projected physical radius, $R_{\rm proj}$ at the distance of each 
cluster. 
Under `source', we refer to the catalogue and plate material used for visual 
inspection of candidates. 
} 
\label{selectable} 
\begin{tabular}{llllllllcccl} 
Cluster & \multicolumn{2}{c}{RA} & \multicolumn{2}{c}{Dec} & $cz_{\rm nom}$ & \multicolumn{1}{c}{Search radius} & \multicolumn{1}{c}{$R_{\rm proj}$} & \multicolumn{1}{c}{magnitude} & \multicolumn{1}{c}{Source} \\ 
& \multicolumn{2}{c}{(B1950)} & \multicolumn{2}{c}{(B1950)} & \kms\ & 
& \hmpc\ & & & & \\ 
& & & & & & & & & & \\ 
7S21 & 00 & 18.6 & +22 & 05 & 5500 & 1$^{\circ}$ & 1.0 & $B \sim 16$ & GSC + POSS II \\ 
Pisces & 01 & 04.5 & +32 & 10 & 4700 & 2$^{\circ}$ & 1.6 & $B = 16$ & APM + POSS I \\ 
HMS0122+3305 & 01 & 20.5 & +35 & 10 & 4600 & 2$^{\circ}$ & 1.6 & $B = 16$ & APM + POSS I \\ 
A0262 & 01 & 49.9 & +35 & 54 & 4500 & 2$^{\circ}$ & 1.6 & $B = 16$ & APM + POSS I \\ 
A0347 & 02 & 19.6 & +41 & 25 & 5300 & 1.5$^{\circ}$ & 1.4 & $B \sim 16$ & GSC + POSS II \\ 
J8 & 02 & 26.0 & +23 & 00 & 9800 & 1.5$^{\circ}$ & 2.5 & $B = 17$ & APM + POSS I \\ 
Perseus & 03 & 15.0 & +41 & 00 & 4800 & 1$^{\circ}$ & 0.8 & $B = 17$ & Poulain + 7S \\ 

\end{tabular} 
\end{table*} 

Galaxies were selected in a cone centred on each cluster position. The 
angular radius of each cone was chosen to give a physical radius of 
1.0--2.5\hmpc\ at the cluster, using the distance suggested by the 
cluster redshift in the CMB frame. In Table~\ref{selectable} we 
summarise the selection criteria used in each cluster. 

For Pisces, A0262, HMS0122+3305, and J8, objects were selected from 
APM scans (see Irwin \& McMahon 1992). The images of all objects 
brighter than $B = 16$ ($B = 17$ for the more distant cluster J8) were 
inspected, using Palomar Sky Survey material. An initial inspection 
served to discriminate galaxies from close pairs of stars, merged 
galaxies and plate defects. In merged objects containing one or more 
galaxy, the magnitude of each galaxy was estimated by eye, given the 
total magnitude of the system. All galaxies were examined and 
morphological types were assigned. Only E and S0 galaxies without 
prominent disks were retained in the final sample. The remaining 
galaxies were cross-referenced with known objects at similar 
positions, using NED\footnote{NED, the NASA/IPAC extragalactic 
database, is operated for NASA by the Jet Propulsion Laboratory at 
Caltech.}. Those with literature redshifts different by more than 
2000 \kms\ from the nominal cluster redshift were deleted from the 
sample. 

For 7S21 and A0347, APM scans were not available at the time of 
selection. The HST Guide Star Catalogue was used to select 
non-stellar objects in these clusters. Suitable candidates were then 
selected and typed by inspection of sky survey plates, and cross 
referenced with NED. 

For galaxies in the Perseus cluster, which lies at low galactic 
latitude, reliable E and S0 galaxies were selected from the work of 
Poulain, Nieto \& Davoust (1992). A few extra ellipticals were added 
from the 7S sample. 

For A2199 and A2634, galaxies were selected from Lucey et al. (1991a). 

For most of the galaxies for which data is presented here, reliable 
positions are available through NED. Cross references are provided, 
with our data, to a reference number from well known catalogues (NGC, 
IC, UGC, CGCG) or from more specialist papers: 
Chincarini \& Rood (1971, CR); 
Bucknell, Godwin \& Peach (1979,BGP); 
Dressler (1980); 
Faber et al. (1989); 
Lucey et al. (1991a); 
Wegner et al. (1996). 
In 
Table~\ref{postable}, we list positions for the galaxies not included 
in the above lists. 

As in most programmes of peculiar velocity measurement, 
the selection criteria described here are somewhat inhomogeneous in 
terms of limiting magnitudes. 
This non-uniformity would result in biases in the cluster distances 
if not handled correctly. Methods for deriving unbiased FP/\dnsig\ 
relations and distances will be discussed and applied in Paper II. 

Note also that morphological selection from sky survey plates is 
necessarily subjective. Andreon (1994) has reported that, for 
galaxies in the Poulain et al. sample, around a half of those 
classified as E by visual inspection of survey plates have Poulain et 
al. types S0 or later. 

\begin{table} 
\caption{Positions for uncatalogued galaxies in the PP sample. 
For all other galaxies studied here, positions are available through 
NED.} \label{postable} 
\begin{tabular}{llcc} 

Cluster & Our name & RA (B1950) & Dec (B1950) \\ 
\\ 
7S21 & S06 & 00 18 44.8 & +21 42 22 \\ 
\\ 
Pisces & Z17005 & 00 56 43.0 & +32 52 04 \\ 
& Z16012 & 00 59 04.2 & +33 20 51 \\ 
& Z01047 & 01 04 12.4 & +32 02 30 \\ 
& Z01032 & 01 05 27.0 & +32 11 13 \\ 
& Z04035 & 01 05 43.7 & +33 06 58 \\ 
& Z10020 & 01 09 05.1 & +31 17 37 \\ 
\\ 
HMS0122+3305 & H01027 & 01 21 00.9 & +33 19 29 \\ 
\\ 
A0262 & A14050 & 01 47 18.8 & +35 58 52 \\ 
& A01094 & 01 47 26.5 & +35 44 09 \\ 
& A01076 & 01 49 36.2 & +35 52 08 \\ 
\\ 
A0347 & B03C & 02 20 01.9 & +42 45 54 \\ 
\\ 
J8 & J07038 & 02 24 03.4 & +23 24 06 \\ 
& J09035 & 02 24 41.2 & +21 45 40 \\ 
& J08035 & 02 24 41.4 & +22 51 39 \\ 
& J01065 & 02 25 49.1 & +22 47 23 \\ 
& J03049 & 02 26 52.2 & +23 44 03 \\ 
& J01055 & 02 26 59.2 & +22 53 12 \\ 
& J01080 & 02 27 46.0 & +22 29 54 \\ 
\end{tabular} 
\end{table}

\section{Spectroscopy} \label{spectroscopy} 

\subsection{Observations} 

Spectroscopic observations were made using the 2.5m Isaac Newton 
Telescope (INT) on La Palma, in 1993 and 1994. Different detectors 
were used in each run : an EEV CCD in 1993, and the faster TEK CCD in 
1994. An EEV chip was used for one night of the 1994 run, due to 
technical problems. This resulted in three spectroscopic datasets 
(hereafter denoted EEV93, EEV94, TEK94), which were each treated 
separately during the course of the data reduction. Instrumental 
details for the three datasets are summarised in 
Table~\ref{instruments}. 

\begin{table*} 
\begin{minipage}{115mm} 
\caption{Spectroscopic instrumentation.} \label{instruments} 
\begin{tabular} {lrrr} 
Dataset & EEV93 & EEV94 & TEK94 \\ 
& & & \\ 
Dates & Nov. 15--22, 1993 & Sep. 6, 1994 & Sep. 3--5 \&\ 7--9, 1994 \\ 
Observers & JRL, MJH, JS & JRL, JS & JRL, JS \\ 
Telescope & 2.5m INT & 2.5m INT & 2.5m INT \\ 
Spectrograph & IDS & IDS & IDS \\ 
Wavelength Range & 4760--5784\AA\ & 4760--5784\AA\ & 4760--5784\AA\ \\ 
Slit size & 3 arcsec & 3 arcsec & 3 arcsec \\ 
CCD & EEV & EEV & TEK \\ 
CCD Dimensions & 1242$\times$1152 & 1242$\times$1152 & 1024$\times$1024 \\ 
Effective aperture & 3.0$\times$3.3 arcsec & 3.0$\times$3.3 arcsec & 3.0$\times$3.5 arcsec \\ 
Number of Galaxy Spectra & 105 & 16 & 211 \\ 
Mean seeing & 1.5 arcsec & 1.5 arcsec & 1.2 arcsec \\ 
\end{tabular} 
\end{minipage} 
\end{table*} 

\subsection{Derivation of spectroscopic parameters} 

Initial reduction of the CCD frames involved bias and dark current 
subtraction, the removal of pixel-to-pixel sensitivity variations 
(using flat field exposures provided by a tungsten calibration lamp) 
and correction for vignetting along the slit (using twilight sky-line 
exposures). 

The spectra obtained covered $\sim$1000\AA\ centred on the Mgb 
triplet, and were sampled with a resolution of $\sim$4\AA\ FWHM. 

Wavelength calibration was performed using arc--lamp exposures, taken 
regularly in the course of the observations, and always after movement 
from one cluster or region to another. A cubic fit between pixel 
number and wavelength for $\sim$18 arc lines gave a maximum rms 
calibration error of $\sim$0.1\AA. 

Spectra were extracted from the frames by simple co-addition of the 
central 5 rows of the galaxy. The resulting effective aperture size 
is tabulated for each dataset in Table~\ref{instruments}. After 
application of a median--filter to remove cosmic ray events, the 
darkest rows on the frame were used to produce a sky spectrum. 

For some galaxies in the EEV93 dataset, sufficient signal-to-noise 
could be obtained only by co-adding spectra resulting from two 
separate exposures. In almost all of these cases, the two exposures 
were taken in immediate subsequence, ensuring the validity of the 
co-addition. 

Cosmic ray events in the galaxy spectra were removed by a combination 
of automatic procedures before extraction, and interactive methods 
applied at the one-dimensional spectrum stage. Features in the 
spectrum resulting from noise in the subtraction of sky--line features 
(especially at 5577\AA) were similarly removed after extraction. 

On each run, spectra were obtained for several G8 to K3 giant stars, 
for use as template spectra. These stars were trailed across the slit 
at a shallow angle during the exposure, to produce an extended 
illumination. Subsequent weighting of these frames, by a typical 
galaxy profile, effects a simulated observation of a galaxy with zero 
velocity dispersion. The extension of illumination has the effect of 
broadening the stellar spectra by $\sim 30$ \kms. 

The method used for measurement of the velocity dispersion, $\sigma$, 
for each galaxy, is based upon the well-known Fourier Quotient method 
of Sargent et al. (1977). In preparation for the application of this 
procedure, continuum levels were subtracted from both the template 
spectrum and the galaxy spectrum, and both were submitted to a cosine 
bell modulation to fix the spectrum ends to zero. The latter step is 
necessary to avoid unphysical signals appearing at all frequencies in 
the Fourier Transforms. 

The method requires also the removal from the spectra of signals 
resulting from noise, inadequate continuum removal and the application 
of the cosine bell. Firstly, a cut is made at high frequencies, to 
remove noise. The results of this method seem to be fairly insensitive 
to the exact value, $k_{\rm{high}}$, chosen for the high frequency 
cut. $k_{\rm{high}} = 200\ \approx(5$\AA$)^{-1}$ has been used 
throughout. Furthermore, a low frequency filter must be applied to 
remove residual continuum features, and the effects of the cosine-bell 
modulation function described above. With the low-frequency cut, 
however, results are found to exhibit a clear trend : velocity 
dispersions are measured to be smaller when $k_{\rm{low}}$ is higher. 
One must choose the cutoff frequency with care. The highest sensible 
$k_{\rm{low}}$ is that which would preserve spectral features in 
spectra of velocity dispersion $\leq500$ \kms. This is $k_{\rm{low}} = 
9$ $\approx(110{\rm \AA})^{-1}$ for our spectra. The lowest sensible 
$k_{\rm{low}}$ is that which is necessary to remove the signal of the 
cosine-bell modulation. This is $k_{\rm{low}} = 6\ 
\approx(170$\AA$)^{-1}$ for our spectra. The portion of the $\sigma - 
k_{\rm{low}}$ plot between these sensible limits is flat to $\sim$5 
per cent\ for most galaxies. 

After discarding a few template spectra which gave consistently 
discrepant results, the velocity dispersions were averaged over 13 
template spectra of 6 different stars, and over values $k_{\rm{low}} = 
6,7,8,9$ adopted for the low frequency filter. 

The uncertainty on each velocity dispersion was quantified by 
repeatedly conducting the measurement after bootstrap resampling of 
the spectrum. This provides an estimate of the random, Poisson-noise 
error on $\sigma$. 

Recession velocities ($cz$) were obtained simultaneously with velocity 
dispersions, as a result of the Fourier Quotient fit. 

The \mg\ line strength index for the magnesium feature was also 
derived for each spectrum. In order to calculate this index, 
independent of the shape of the instrumental response curve, the 
spectra were first flux-calibrated by reference to spectrophotometric 
standard stars observed during the runs. For certain observations, no 
appropriate flux-standard was obtained, so a few galaxies have no \mg\ 
measurement. Initial flux calibration of the EEV93 data was found to 
be unsatisfactory, due to a strong gradient in chip response across the 
spectral region being used. The calibration was improved by an extra 
step in which we derived the response curve of the EEV relative to the 
TEK, using a star common to both datasets, before calibrating to the 
absolute standard of flux. A similar problem for the EEV94 data could 
not be resolved in this manner, since there are no stars in common 
between that dataset and the TEK94 data. As a result there are no \mg\ 
measurements from the EEV94 observations. Uncertainties in the \mg\ 
indices were calculated simply from the noise characteristics of the 
chip employed. 

\subsection{Raw spectroscopic data and internal comparisons} 

\begin{figure} 
\epsfig{file=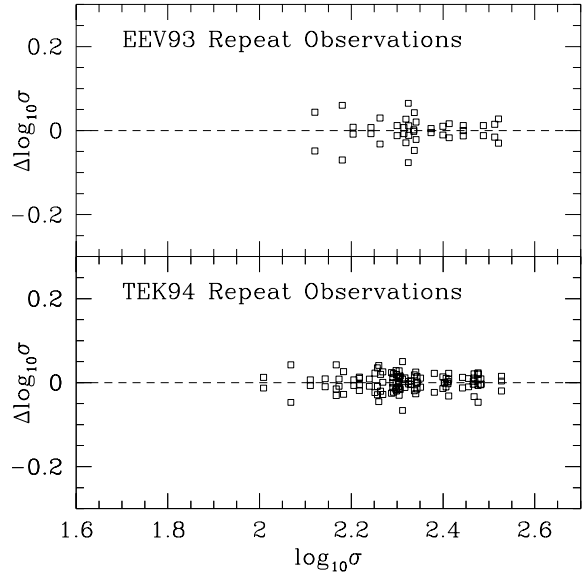,height=80mm,width=80mm} 
\caption{Scatter of repeat velocity dispersion measurements 
within the datasets presented here. In each panel, the 
horizontal axis 
is the mean quantity derived from the dataset; the 
vertical axis 
is the 
deviation of each individual measurement from that mean. Note that 
there are no internal repeats within the EEV94 dataset. For ease of 
comparison, the axis limits for this plot are the same as for the 
equivalent plot in Davies et al. (1987)} \label{sigint} 
\end{figure}

\begin{figure} 
\epsfig{file=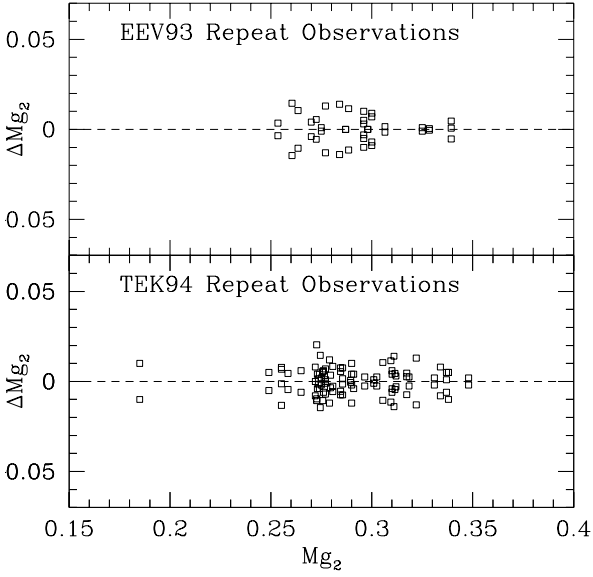,height=80mm,width=80mm} 
\caption{As for Figure~\ref{sigint}, but for \mg\ index measurements.} 
\label{mgint} 
\end{figure}

Table~\ref{rawspectable} presents the raw spectroscopic data obtained, 
including formal errors. Over half of the galaxies were observed more 
than once. 
Comparisons between repeat measurements in the two large datasets 
(EEV93 and TEK94) are illustrated in 
Figures~\ref{sigint} and~\ref{mgint} for velocity dispersion and \mg\ 
index, respectively. Note that there are no repeat observations within 
the small EEV94 dataset. The implied observational errors in each 
dataset are summarised in Table~\ref{intratab}. Weighting the 
$\sigma$ uncertainties in TEK94 and EEV93 by the number of 
observations in each dataset, we obtain a typical measurement error of 
0.032 dex per measurement.

For comparison, the 7S Lick data exhibit an internal uncertainty of 
0.057 dex in $\sigma$. The higher quality 7S velocity dispersions are 
accurate to 0.036 dex (Davies et al.). The mean Poisson error on 
$\sigma$ is 0.023 dex (TEK94) and 0.029 dex (EEV93). Non-Poissonian 
effects therefore account for an appreciable portion of the observed 
scatter, especially for the earlier dataset. 

Uncertainties on the \mg\ measurements are typically 0.010 mag., and 
are fully accounted for by the mean photon-noise error.

\begin{table} 
\caption{ 
Uncertainties in the EEV93 and TEK94 datasets, as judged from the 
scatter of repeat measurements. For each parameter, $N$ indicates 
the number of galaxies for which comparisons could be made.} 
\label{intratab} 
\begin{tabular}{lcccccc} 
Dataset & $N$ & rms & $N$ & rms & $N$ & rms \\ 
& & $\Delta(\log\sigma)$ & & $\Delta$(\mg) & & $\Delta(cz)$ \\ 
\\ 
TEK94 & 48 & 0.027 & 44 & 0.010 & 48 & 31.2 \\ 
\\ 
EEV93 & 20 & 0.041 & 20 & 0.011 & 20 & 23.1 \\ 
\\ 
\end{tabular} 
\end{table} 

\subsection{The aperture correction} 

The physical size of that central part of a galaxy, observed through a 
fixed aperture, is larger for a more distant galaxy than for one 
nearby. Since galaxies, in the mean, exhibit a negative radial 
gradient in both $\log\sigma$ and \mg, a correction must be applied to 
the raw data before use. Furthermore, to compare measurements made 
using different aperture sizes, a similar correction is clearly 
necessary. J\o rgensen, Franx \& Kj\ae rgaard \shortcite{jfkspec} 
present an analysis based on the observed radial gradients in 
$\log\sigma$ and \mg\ for nearby galaxies. They find that a power law 
provides an adequate description of the required correction: 
\begin{equation} \label{genapcorr} 
\log\frac{\sigma_{\rm{corr}}}{\sigma_{\rm{obs}}} = 0.04\log\frac{r_{\rm{ap}}}{r_{\rm{norm}}} 
\end{equation} 
where $r_{\rm{ap}}$ is the physical radius sampled by that circular 
aperture from which one obtains the same $\sigma_{\rm{obs}}$ as 
through the actual aperture used. For a rectangular aperture of 
angular dimensions $x$ and $y$ (in radians), and a galaxy at distance 
$d$, the equivalent aperture is 
\begin{equation} 
r_{\rm{ap}} \approx 1.025(\frac{xy}{\pi})^{1/2} d 
\end{equation} 
where the correction factor 1.025 is included to provide an improved 
match to more detailed models. 
An independent analysis, based on measured velocity dispersion 
profiles, supports the size of this correction. 

For the normalisation, we follow J\o rgensen et al. in adopting a 
physical diameter $2r_{\rm{norm}}$ of 1.19\hkpc. This is equivalent to 
an angular diameter of 3.4 arcsec for Coma cluster galaxies.

J\o rgensen et al. find the average radial gradient of the \mg\ index 
to be so similar to that of the velocity dispersion, that 
equation~\ref{genapcorr} may be used for the \mg\ aperture correction, 
with a simple substitution of \mg\ for $\log\sigma$.

\subsection{Matching of spectroscopic datasets onto a new `standard system'} 

\begin{table} 
\caption{Run-to-run comparisons of spectroscopic data. $N$ indicates the number 
of galaxies involved in each comparison.} \label{specruntoruntab} 
\begin{tabular}{lrrr} 
Comparison & \multicolumn{1}{c}{$N$} & \multicolumn{1}{c}{Mean $\Delta(\log\sigma$)} & \multicolumn{1}{c}{Dispersion} \\ 
\\ 
EEV93 -- TEK94 & 46 & --0.009$\pm$0.006 & 0.042 \\ 
EEV94 -- TEK94 & 10 & 0.014$\pm$0.012 & 0.039 \\ 
\\ 
Comparison & \multicolumn{1}{c}{$N$} & \multicolumn{1}{c}{Mean $\Delta($\mg)} & \multicolumn{1}{c}{Dispersion} \\ 
\\ 
EEV93 -- TEK94 & 46 & --0.010$\pm$0.002 & 0.013 \\ 
\end{tabular} 
\end{table} 

\begin{figure} 
\epsfig{file=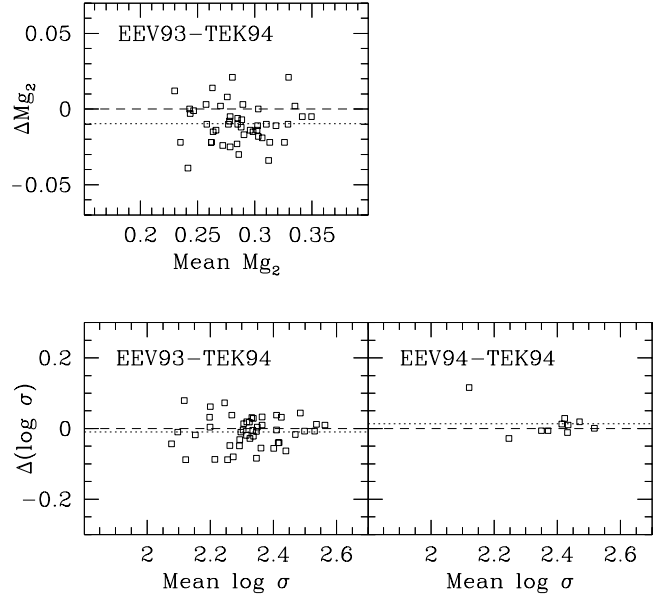, height=90mm, width=90mm} 
\caption{ 
Comparison between spectroscopic parameters derived from the three 
datasets presented in this paper. The data are aperture corrected, 
but nevertheless exhibit systematic offsets from run to run, as 
shown by the dotted lines. The offsets are quantified in 
Table~\ref{specruntoruntab}.} 
\label{specruntorunfig} 
\end{figure}

\begin{figure*} 
\epsfig{file=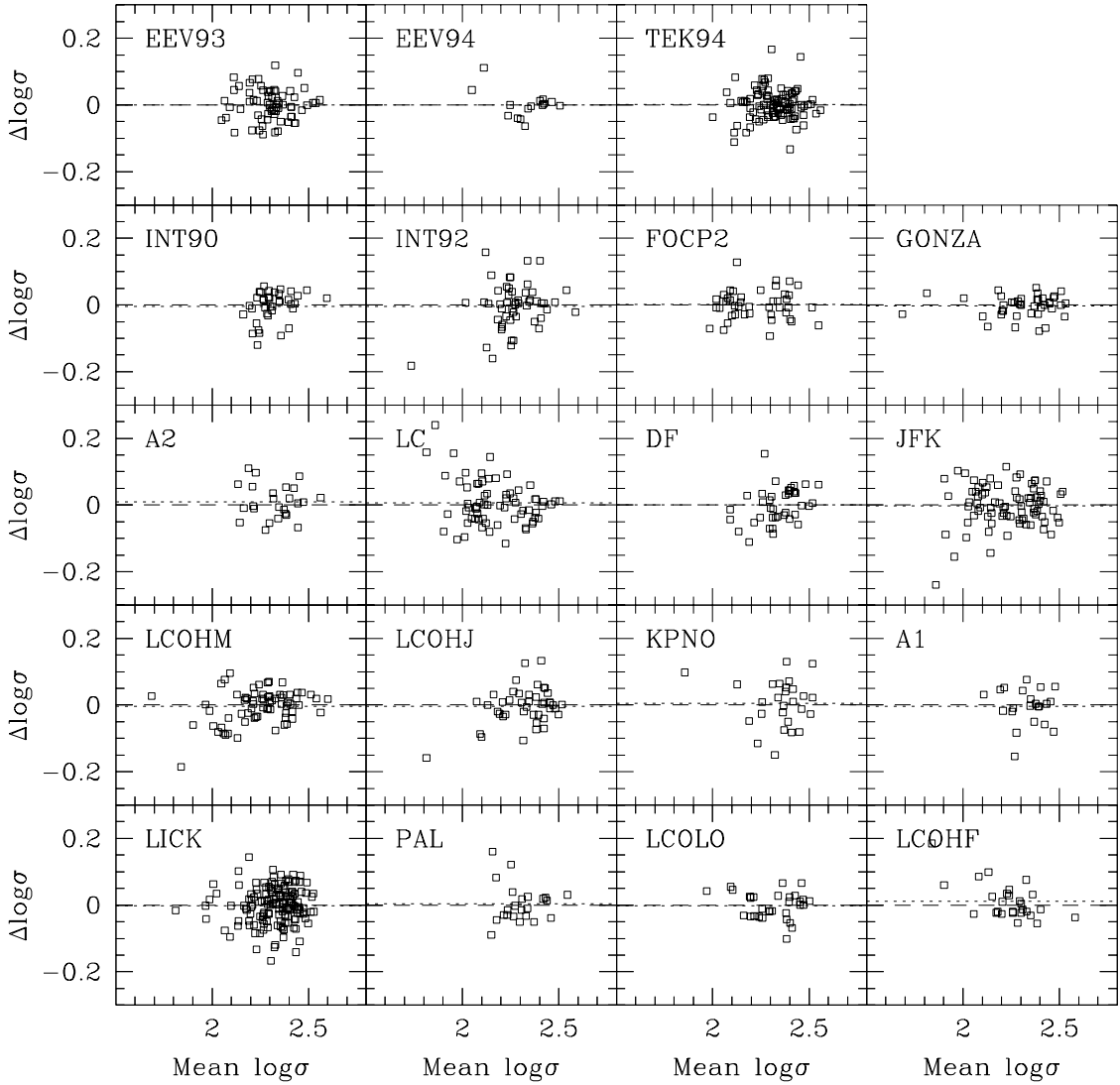, height=160mm, width=160mm} 
\caption{ 
Consistency of the merged system of velocity dispersion 
measurements. For the each galaxy in the TEK94 panel, we compute 
the mean (fully corrected) TEK94 measurement, and the mean (fully 
corrected) value using all the other data -- excluding TEK94. We 
plot as $\Delta\log\sigma$ the difference between the `TEK94-only' 
and the `all-but-TEK94' values. All 19 velocity dispersion systems 
are treated in this way. Note that we include in these plots all 
measurements, including those which were not used in the derivation 
of the corrections. The small offsets still present in some plots 
(indicated by dotted lines) are a result of these outlying points 
and low $\sigma$ galaxies.} 
\label{sysmatsigs} 
\end{figure*} 

\begin{figure*} 
\epsfig{file=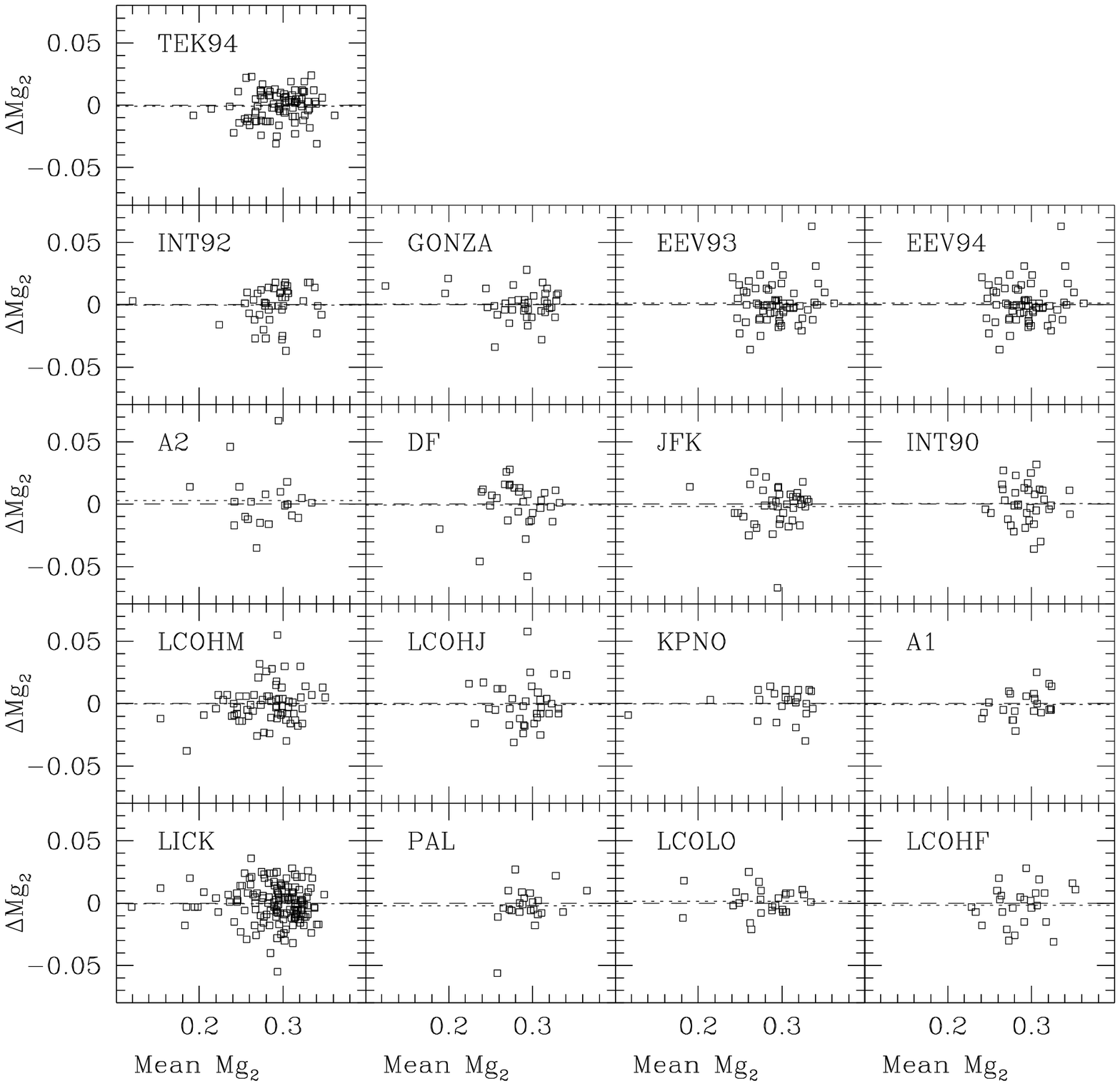, height=160mm, width=160mm} 
\caption{As for figure~\ref{sysmatsigs}, but for the 17 systems of \mg\ 
measurements.} \label{sysmatmgs} 
\end{figure*}

In order to construct large samples of peculiar velocity data, we 
require that velocity dispersions measured at different telescopes 
match as accurately as possible. At the PP distance, a one per cent 
systematic error in $\sigma$ corresponds to 50 \kms\ in peculiar 
velocity. A systematic difference between the velocity dispersions 
measured on telescopes in opposite hemispheres would thus generate a 
spurious bulk flow. Despite careful attempts to correct the velocity 
dispersions for aperture effects, systematic differences between 
velocity dispersions measured from different datasets persist at the 
$\sim$3 per cent level. Such offsets are present even between the 
three datasets presented here (as illustrated in 
Figure~\ref{specruntorunfig} and Table~\ref{specruntoruntab}), despite 
the use of very similar observational methods and data reduction 
techniques. 

The removal of systematic offsets can be achieved by intercomparison 
of results for galaxies common to two or more systems. To this end, 
our data include many galaxies observed to improve overlap with 
existing systems. In this section, we consider velocity dispersion and 
\mg\ data on 19 and 16 different systems, respectively. In order to 
take account of zero-point differences reported by Dressler (1984), 
the 7S LCOHI data have been subdivided into the three constituent runs 
from which they derive. 

Many galaxies have measurements on more than two systems. Therefore 
in order to determine self-consistent corrections between different 
systems, a simultaneous fit for all of the offsets is necessary. The 
fit is performed using velocity dispersion and \mg\ data corrected to 
the J\o rgensen et al. (1995b) standard physical aperture size of 1.19 
$h^{-1}$ kpc. We determine the corrections necessary to bring all 
systems into the best possible agreement with each other. We adopt 
the fully-corrected Lick system (Davies et al.\ 1987) as the standard 
and determine the remaining corrections as follows. 
Let $s = \log_{10}(\sigma)$ and 
let $i$,$j$ and $k$ index the measurement, galaxy and system 
respectively. We obtain the 
corrections $\Delta_k$, needed to bring each system into agreement with 
Lick, 
by minimising a $\chi^2$ statistic 
\begin{equation} 
\chi^2 = \sum_i \frac{(s_i + \Delta_k - \overline{s}_j)^2}{e_k^2} 
\end{equation} 
where $e_k$ is the error in $s_i$ (assumed to be the same for all 
galaxies in a given system) and $\overline{s}_j$ is the error-weighted 
mean of all corrected measurements of the same galaxy. 

We determine the errors $e_k$ for each system by adjusting these so 
that the reduced $\chi^2$ is unity, both when the system is included 
and when it is excluded from the comparisons. This external error 
($e_{\rm ext}$) is typically 10 -- 25 per cent larger than the 
internal error ($e_{\rm int}$) estimated from repeat measurements on 
the same system. 

The overlap data set of velocity dispersion measurements (galaxies 
with velocity dispersions on more than one system) consists of 1281 
measurements for 350 different galaxies. We exclude galaxies with 
$\overline{s} < 2$ as these may be subject to large random and 
systematic errors (J\o rgensen et al. 1995b). We also exclude 
individual velocity dispersion measurements which are inconsistent at 
the $3.5 \sigma$ level with the other measured velocity dispersions of 
the same galaxy. The velocity dispersions so excluded are A2634-F1201 
(EEV93 $s=2.0784$), A1656D-136 (INT90 $s=2.0888$), N386 (KPNO 
$s=1.7923$), N548 (LICK $s=1.8856$) and VELA-G22 (FOCP2 $s=1.9237$). 

The overlap data set of \mg\ measurements (galaxies with measurements 
on more than one system) consists of 1013 measurements of 270 
different galaxies on 16 systems (the LC, FOCP2 and EEV94 systems have 
no \mg\ data) . In addition to the galaxies excluded in the velocity 
dispersion comparison, we also exclude the following data which are 
inconsistent with other measurements of the same galaxy at the $3.5 
\sigma$ level: N1282 (PAL \mg\ = 0.0245), N1549 (A2 \mg\ = 0.342 and 
JFK \mg\ = 0.264) N4564 (EEV93 \mg\ = 0.350) and N6702 (GONZA \mg\ = 
0.243 and TEK94 \mg\ = 0.288). 

\begin{table} 

\caption{Corrections required to bring each system of (aperture-corrected) 
velocity dispersion measurements into agreement with the standard system. 
$e_{\Delta}$ are the errors on each correction. $N_{ov}$ represents, for each 
system, the number of galaxies in the overlap dataset, ie having measurements 
on other systems.} \label{sigcorrs} 

\begin{tabular}{lrrrrrr} 
\multicolumn{1}{c}{Name} & 
\multicolumn{1}{c}{Source} & 
\multicolumn{1}{c}{N} & 
\multicolumn{1}{c}{$e_{\rm int}$} & 
\multicolumn{1}{c}{$e_{\rm ext}$} & 
\multicolumn{1}{c}{$\Delta$} & 
\multicolumn{1}{c}{$e_{\Delta}$} \\ 
LICK & 1 & 276 & 0.052 & 0.055 & $\equiv 0$ & $\equiv 0$ \\ 
PAL & 2 & 23 & -- & 0.045 & --0.0241 & 0.0116 \\ 
LCOLO & 1 & 61 & 0.039 & 0.040 & 0.0115 & 0.0098 \\ 
LCOHF & 3 & 25 & -- & 0.035 & --0.0067 & 0.0105 \\ 
LCOHM & 3 & 73 & 0.023 & 0.035 & 0.0106 & 0.0072 \\ 
LCOHJ & 3 & 61 & 0.021 & 0.035 & 0.0021 & 0.0086 \\ 
KPNO & 1 & 27 & -- & 0.065 & 0.0142 & 0.0139 \\ 
A1 & 1 & 27 & 0.024 & 0.040 & --0.0057 & 0.0113 \\ 
A2 & 1 & 42 & 0.036 & 0.045 & 0.0176 & 0.0102 \\ 
LC & 4 & 72 & 0.033 & 0.035 & --0.0127 & 0.0096 \\ 
DF & 5 & 41 & -- & 0.044 & --0.0038 & 0.0112 \\ 
JFK & 6 & 76 & -- & 0.040 & 0.0011 & 0.0089 \\ 
INT90 & 7 & 59 & 0.038 & 0.040 & --0.0177 & 0.0094 \\ 
INT92 & 8 & 60 & 0.040 & 0.055 & 0.0080 & 0.0096 \\ 
FOCP2 & 9 & 67 & 0.034 & 0.035 & --0.0063 & 0.0094 \\ 
GONZA & 10 & 38 & -- & 0.014 & 0.0222 & 0.0054 \\ 
EEV93 & 11 & 86 & 0.040 & 0.040 & --0.0014 & 0.0082 \\ 
EEV94 & 11 & 15 & -- & 0.040 & --0.0115 & 0.0111 \\ 
TEK94 & 11 & 152 & 0.027 & 0.030 & --0.0063 & 0.0059 \\ 
\\ 
\end{tabular} 

Sources: \\ 
\\ 
1 -- Davies et al.\ (1987)\\ 
2 -- Davies et al.\ (1987) -- Palomar observations wrongly attributed to LCOHI dataset (see Dressler et al. 1987)\\ 
3 -- Davies et al. (1987) LCOHI data subdivided according to run: Feb. 
82 (LCOHF); Mar. 83 (LCOHM) and Jan. 84 (LCOHJ)\\ 
4 -- Lucey \& Carter (1988)\\ 
5 -- Dressler, Faber \& Burstein (1991)\\ 
6 -- J\o rgensen, Franx \& Kj\ae rgaard (1995b)\\ 
7 -- Lucey, Guzman, Carter \& Terlevich (1991)\\ 
8 -- Lucey, Guzman, Steel \& Carter (1997)\\ 
9 -- Lucey et al. (1998)\\ 
10 -- Gonzales (1993)\\ 
11 -- This paper \\ 
\end{table} 

\begin{table} 
\caption{As for Table~\ref{sigcorrs}, but for \mg\ measurements.} \label{mgcorrs} 

\begin{tabular}{lrrrrrr} 
\multicolumn{1}{c}{Name} & 
\multicolumn{1}{c}{Source} & 
\multicolumn{1}{c}{N} & 
\multicolumn{1}{c}{$e_{\rm int}$} & 
\multicolumn{1}{c}{$e_{\rm ext}$} & 
\multicolumn{1}{c}{$\Delta$} & 
\multicolumn{1}{c}{$e_{\Delta}$} \\ 
LICK & 1 & 274 & 0.008 & 0.011 & $\equiv 0$ & $\equiv 0$ \\ 
PAL & 2 & 22 & -- & 0.014 & --0.0143 & 0.0026 \\ 
LCOLO & 1 & 53 & 0.011 & 0.010 & --0.0032 & 0.0024 \\ 
LCOHF & 3 & 24 & -- & 0.013 & --0.0139 & 0.0046 \\ 
LCOHM & 3 & 68 & 0.004 & 0.013 & --0.0086 & 0.0023 \\ 
LCOHJ & 3 & 53 & 0.007 & 0.013 & --0.0185 & 0.0029 \\ 
KPNO & 1 & 24 & -- & 0.011 & --0.0034 & 0.0028 \\ 
A1 & 1 & 27 & 0.012 & 0.009 & 0.0074 & 0.0029 \\ 
A2 & 1 & 33 & 0.005 & 0.010 & --0.0132 & 0.0034 \\ 
DF & 5 & 31 & -- & 0.017 & --0.0035 & 0.0040 \\ 
JFK & 6 & 40 & -- & 0.011 & --0.0017 & 0.0024 \\ 
INT90 & 7 & 54 & 0.012 & 0.015 & 0.0061 & 0.0030 \\ 
INT92 & 8 & 51 & 0.013 & 0.012 & 0.0168 & 0.0027 \\ 
GONZA & 10 & 37 & -- & 0.007 & --0.0048 & 0.0017 \\ 
EEV93 & 11 & 83 & 0.010 & 0.012 & 0.0172 & 0.0021 \\ 
TEK94 & 11 & 139 & 0.009 & 0.009 & 0.0071 & 0.0016 \\ 
\\ 
\end{tabular} 
Sources as in Table~\ref{sigcorrs}.\\ 
\end{table} 

Tables~\ref{sigcorrs} and~\ref{mgcorrs} summarise the required 
corrections to velocity dispersion and \mg, respectively. Note that, 
because of the interdependencies between the different corrections, 
the simple pair offsets of Table~\ref{specruntoruntab} are not 
trivially related to those derived here by simultaneous fits. In 
Figures~\ref{sysmatsigs} and~\ref{sysmatmgs} we illustrate the level 
to which systematic offsets are removed by the application of the 
derived corrections. 

The errors are determined by bootstrap resampling the master data file 
and computing the corrections from the resampled file. This procedure 
allow us to determine not only the error on the correction to each 
system but also the correlation between the corrections for different 
systems. Using the bootstrap values of these corrections, we can 
generate mock merged data sets and so determine for a given cluster 
the error in the mean correction. This is an estimate of the mean 
systematic error in $s$, which will generally depend on the systems 
merged for the cluster, their relative proportions and their 
covariance. For the PP sample, we find that for all clusters this 
error is $\sim$1.5 per cent in $\sigma$. This translates to a 
systematic error of $\sim$2 per cent in distance, or $\sim$100\kms\ at 
PP. 

\subsection{Correction and combination of spectroscopic data} 

In this section, we briefly summarise the recipe for converting the 
raw spectroscopic data tables into the corrected and combined 
measurements to be used in the peculiar velocity analyses. 

In order to combine multiple $\sigma$ and \mg\ observations for a 
galaxy, it is first necessary to ensure that all the sources of data 
are on a consistent system. To this end we correct the EEV93, EEV94 
and TEK94 systems for aperture effects, and scale them onto our new 
`standard system' using the offsets listed in Tables~\ref{sigcorrs} 
and~\ref{mgcorrs}. The distance used in calculating the aperture 
correction is the median redshift of the relevant cluster, or (if not 
part of the cluster sample) the individual galaxy redshift. 

The data for multiply-observed galaxies are then combined to give a 
weighted mean $\log\sigma$, and weighted mean \mg. The weight of each 
measurement is assigned according to the external error on the dataset 
from which it derives. In constructing the means, we exclude the 
($>3.5\sigma$) deviant measurements as flagged above. 

It should be stressed that the external datasets (LICK, FOCP2, etc.) 
are used only to derive the necessary corrections, and to identify 
outlying measurements. The mean parameters are calculated using data 
drawn only from EEV93, EEV94 and TEK94.

Recession velocities are combined by correcting the EEV93 and EEV94 
systems according to their offsets from TEK94, before computing a 
simple mean $cz$. The relative offsets are EEV93 -- TEK94 = $-10 \pm 5 
\kms$ and EEV94 -- TEK94 = $-4 \pm 10 \kms$, derived from 45 and 10 
galaxies respectively. 

\begin{figure} 
\epsfig{file=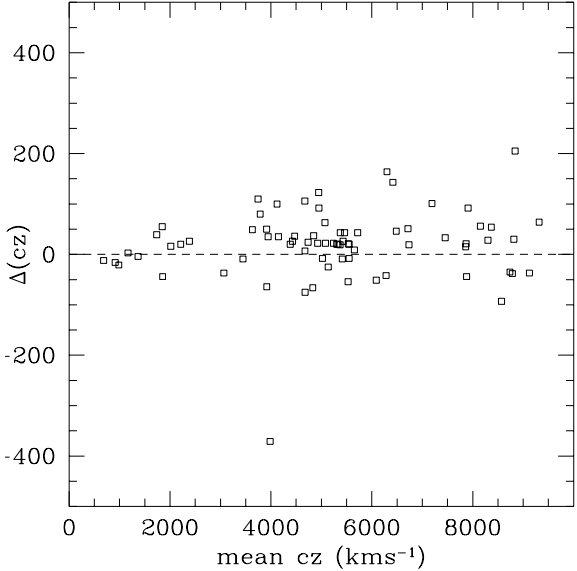,height=80mm,width=80mm} 
\caption{ 
Comparison of our recession velocity measurements with the recession 
velocities adopted by 7S. We plot the difference in $cz$ in the 
sense (`This Study' -- 7S), against the mean of the two 
measurements. External data for 75 galaxies are drawn from Faber et 
al. (1989) and Davies et al. (1987). The discrepant point is for 
N1272.} 
\label{czvs7s} 
\end{figure}

We have compared the resulting mean recession velocity measurements 
with those adopted by 7S (Faber et al. 1989, Davies et al. 1987). The 
median offset is $22\pm 13 \kms$, with our velocities being the 
larger. The comparison is displayed in Figure~\ref{czvs7s}. The most 
discrepant point is galaxy N1272 (P17). For this galaxy, we have seven 
concordant measurements of $cz$. 

Table~\ref{mastable} presents the fully corrected and 
combined spectroscopic data, scaled to the `standard' system, for galaxies in 
the cluster sample. This table includes only those galaxies for which 
complementary photometric data has been obtained.

\section{Photometry} \label{photometry} 

\subsection{Introduction} 

The photometric observations were made in the Kron--Cousins R bandpass. 
For the \dnsig\ relation, we have defined the R-band \dn\ parameter to 
be that diameter which encloses a mean surface brightness 
$\langle\mu\rangle_{R} = 19.23$ mag. arcsec$^{-2}$. If the typical 
(extinction- and $k$-corrected) $V - R$ colour for early-type galaxies 
is 0.57, as indicated by the BVR photometry of Colless et al. (1993), 
then our R-band \dn\ diameters will be well matched to the V-band 
system of Lucey et al. (1991b), and to the B-band work of Burstein et 
al. (1987). At the distance of the clusters studied here, the typical 
\dn\ diameter, so defined, is comfortably large compared to the seeing 
disk, yet not so large that sky subtraction errors become significant. 
The quantities measured for use in the FP distance indicator are the 
effective diameter $A_{\rm e}$, and the mean surface brightness within 
effective diameter, denoted $\langle\mu\rangle_{\rm e}$. 
\subsection{Observations and initial data reduction} 

CCD photometry was obtained on the 1-m Jacobus Kapteyn Telescope 
(JKT) on La Palma in 1993 November and 1994 September. 
Table~\ref{photinst} summarises the instrumental configuration used. 
The observations were made with the RGO `Harris' R filter which, 
in combination with a typical CCD response, provides a close 
match to the standard Kron--Cousins R bandpass. 
The images covered an area of 6.6 $\times$ 6.1 arcmin$^2$, at 
a scale of 0.31 arcsec pixel$^{-1}$. The initial reduction of the 
CCD images followed standard procedures of bias-subtraction 
and flat-fielding, using Starlink software. 
The photometric calibration was achieved by observations of 
Landolt (1983, 1992) standard stars and fields. 
At least 12 Landolt stars/fields were observed each night and 
an online assessment of photometric conditions was employed 
to track the stability of the atmospheric extinction. 
For the calibration mapping we used the equation, 
\begin{equation} 
R = r_{\rm inst} + ZP - k_R X + C (B-V) 
\end{equation} 
where $R$ is Landolt's listed R-band magnitude, 
$B\,-\,V$ is the listed colour, 
$r_{\rm inst}$ is the instrumental magnitude, 
$X$ is the airmass, $ZP$ is the photometric zero-point, 
$k_R$ the R-band extinction per airmass and $C$ is the colour 
term. We solved for the $ZP$, $k_R$ and $C$ terms 
by minimising the residuals. 
Five nights (out of a total 14 allocated) were photometric. 
The residual scatter of the standard stars on these 
nights was less than 0.015 mag. The $k_R$ term was typically 
0.10. The colour term, $C$, was only --0.011, 
confirming the excellent match of the RGO `Harris' R filter 
to the standard Kron--Cousins R system. 
For the limited $B\,-\,V$ colour range of early-type galaxies 
in our study this colour term can be safely included in the 
zero-point term, and observations in R-band alone can be used. 
In order to assess the reliability of our photometric 
measurements and run-to-run variations, a large 
number of our target galaxies were observed more than once 
(see below). FWHM seeing (measured from stellar profiles on the 
target galaxy images) ranged from 0.7 to 3.0 arcsec, with a 
typical value of 1.3 
arcsec. 
\begin{table} 
\caption{Photometric Instrumentation} \label{photinst} 
\begin{tabular}{lrr} 
Dates & Nov. 15--22, 1993 & Sep. 1--7, 1994 \\ 
Observers & JRL, MJH, JS & JRL, JS \\ 
Telescope & 1.0m JKT & 1.0m JKT \\ 
CCD & EEV & EEV \\ 
CCD Dimensions & 1280$\times$1180 & 1280$\times$1180\\ 
Pixel Scale & $0.31$ arcsec & $0.31$ arcsec \\ 
Filter & Harris R & Harris R \\ 
\end{tabular} 
\end{table}

\subsection{Derivation of photometric parameters} 

For each galaxy, circular aperture magnitudes were determined in 
diameter steps of approximately 0.1 dex from 4 arcsec out to $\sim$60 
arcsec. Contaminating stars and galaxies were removed interactively 
from each target galaxy. 
Aperture magnitudes were corrected for galactic extinction and for 
cosmological $k$-dimming. For the R-band extinction, we adopt $A_{R} = 
2.35 E(B-V)$ where $E(B-V)$ are the reddening values of Burstein \& 
Heiles (1984). For the $k$-correction, we use $-1.0 z$ (Oke \& 
Sandage 1968, Frei \& Gunn 1994). A correction for the $(1+z)^{4}$ 
surface brightness dimming is also applied. 

To derive the parameters \dn, $A_{\rm e}$ and $\langle\mu\rangle_{\rm 
e}$, we fit a de Vaucouleurs $R^{1/4}$ profile to the aperture 
photometry. Seeing effects in the aperture magnitudes cannot be 
ignored in this procedure, and are here corrected for by an improved 
version of the method first reported by Bower, Lucey and Ellis (1992). 
Whereas Bower et al. calculate the seeing corrections appropriate for 
a galaxy of true effective radius 5 arcsec, and apply these to all 
galaxies, we have compiled correction tables for a range of true 
radii, and use an iterative technique to select the table required for 
a given galaxy. Convergence to a corrected $A_{\rm e}$ value is very 
rapid. In practice this improved correction scheme leads to 
measurements which are in good agreement with those made using the 
original Bower et al. method. For only five images, out of a total 
245, do we find \dn\ or FP parameters which change by more than 1 per 
cent (distance equivalent) in adopting the new corrections. 

The typical rms residual from the $R^{1/4}$ law fit is 0.02 mag. 
The four worst-fit galaxies have residuals of 0.05--0.09 mag. rms. 
Saglia et al. (1997) have recently investigated the effect of 
fitting a pure $R^{1/4}$ law to galaxies with substantial disk components. 
They show that such a fit to a galaxy with disk-to-bulge ratio 0.2 
can result in $A_{\rm e}$ measurements which are wrong by as much as 
30\%. Whilst this severely affects the determination of 
effective radius and of surface brightness, the combination $\log 
A_{\rm e} - 0.3 \langle\mu\rangle_{\rm e}$ (which enters into the 
Fundamental Plane) is robust against the presence of a disk, since the 
errors on $A\sb{e}$ and $\langle\mu\rangle_{\rm e}$ are correlated. The 
\dn\ parameter, defined by interpolation of the data, rather than from 
a global profile fit, is also insensitive to this effect. 
We note also, that a bias in cluster distances will only result from this 
effect if, from cluster to cluster, substantially different morphological 
proportions are sampled. 

The final fully-corrected photometric parameters are presented in 
Table~\ref{rawphotable}. For comparison with future work, we tabulate 
also the uncorrected R-band magnitude for each galaxy, as measured 
within an aperture of 20 arcsec.

\subsection{Internal comparisons and combination of photometric data} 

\begin{figure} 
\epsfig{file=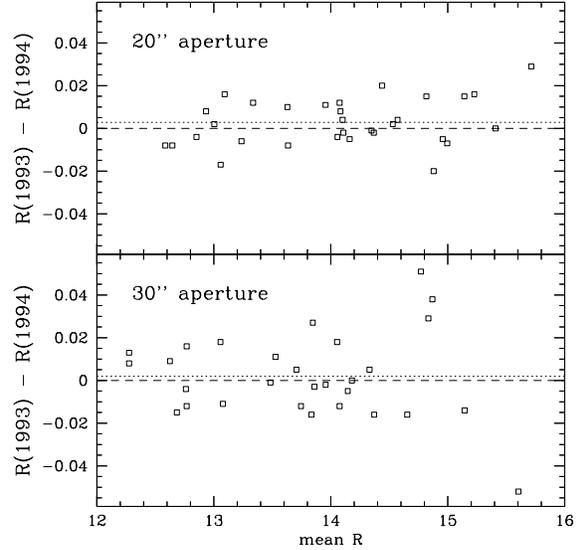,height=80mm, width=80mm} 
\caption{ 
Year to year comparison of galaxy magnitudes, within apertures of 20 
arcsec and 30 arcsec diameter. The dotted line represents the mean 
offset in each case. The scatter in the plot is 0.01 mag. at 20 
arcsec aperture, and 0.02 mag. at 30 arcsec. }\label{apmrpts} 
\end{figure}

To assess the consistency of our photometric system from year-to-year, 
we have compared, for each galaxy in common, the mean derived aperture 
magnitude from the 1993 run, with that from the 1994 data. The 
comparison is shown in Figure~\ref{apmrpts}, for apertures of 20 
arcsec and 30 arcsec diameter. At 20 arcsec, the mean offset is 
0.003$\pm$0.002 mag, and the scatter 0.011 mag. The offset in the 30 
arcsec aperture magnitudes, is 0.002$\pm$0.004 mag, with a scatter of 
0.020 mag. The increased scatter for the larger aperture results from 
the treatment of contaminating sources, companion galaxies, etc. We 
are confident, therefore, that our photometric system is internally 
consistent to better than 0.01 mag. Applying the same year-to-year 
test for \dn\ measurements, we find an offset between the runs of 
0.000$\pm$0.001 dex. 

Since our photometric data are on the same system, we can combine 
repeated measurements of log\dn, $\log A_{\rm e}$ and 
$\langle\mu\rangle_{\rm e}$, to give simple mean values. These are 
presented in Table~\ref{mastable} along with the spectroscopic 
parameters for each galaxy. 

\begin{figure} 
\epsfig{file=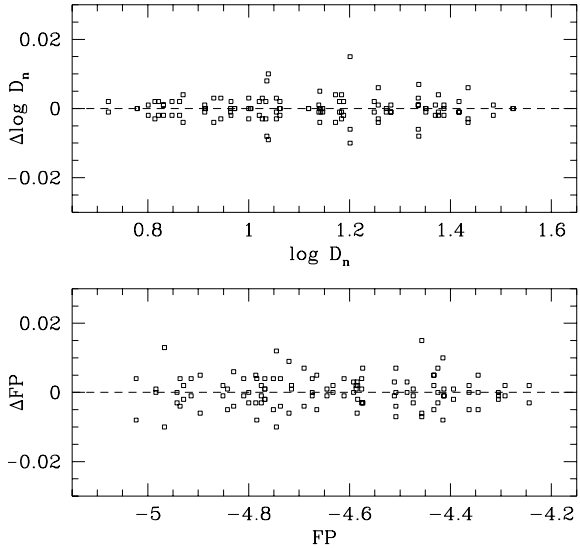,height=80mm, width=80mm} 
\caption{ 
Upper panel : Scatter of repeat \dn\ measurements. For each galaxy with more than one observation, the individual measurements of log \dn\ are plotted against the mean value. 
Lower panel : Scatter of repeat measurements of the Fundamental Plane 
combination $\log A_{\rm e} - 0.3\langle\mu\rangle_{\rm e}$, 
denoted FP. (Note however, that the coefficient of $\langle\mu\rangle_{\rm e}$ 
may not be precisely the same as for the FP distance indicator adopted in 
Paper II). 
} 
\label{dnrpts} 
\end{figure}

From a subset of 50 galaxies which have repeat observations, an 
estimate can be made of the typical uncertainty in our measurements of 
the photometric parameters. Figure~\ref{dnrpts} shows the comparison 
of these measurements. The scatter implies an error of 0.005 in each 
determination of log\dn. For the FP parameters taken individually, 
the scatters are larger: 0.032 dex on $A_{\rm e}$ and 0.113 mag. 
arcsec$^{-2}$ on $\langle\mu\rangle_{\rm e}$. The errors on these 
parameters are correlated, however. If we construct the quantity $\log 
A_{\rm e} - 0.3 \langle\mu\rangle_{\rm e}$, the combination often used 
to give an edge-on projection of the Fundamental Plane, we find that 
the uncertainty in this quantity is only 0.006, only slightly larger 
than that on log\dn. 

\subsection{External comparisons} 

\subsubsection{Aperture photometry} 

\begin{figure} 
\epsfig{file=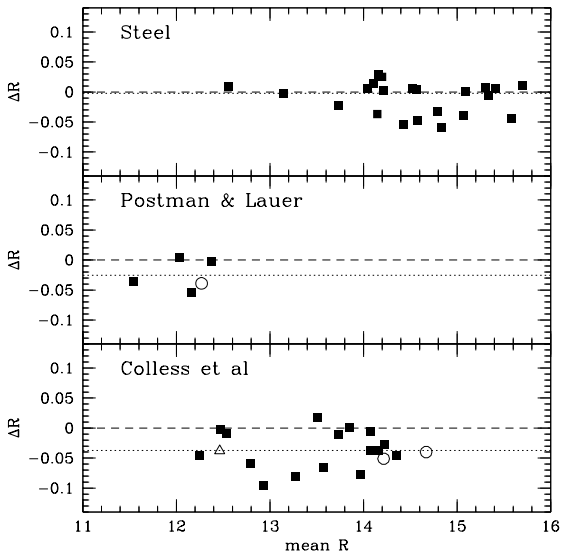,height=80mm, width=80mm} 
\caption{ 
Comparison of our CCD aperture photometry with data from the 
literature. Upper panel : comparison with Steel (1997), at an 
aperture of 20 arcsec. Middle panel : comparison with Postman and 
Lauer (1995). The data used are for apertures of 63 arcsec (filled 
squares) and 79 arcsec (open circles). Lower panel : comparison 
with Colless et al. (1993), The comparison is performed for 
magnitudes within 19.2 arcsec (open circles), 29.9 arcsec (filled 
squares), 39.5 arcsec (open triangle). The comparisons are plotted 
in the sense `this work' -- `literature'. The dotted line indicates 
the mean offset in each case.}\label{apext} 
\end{figure}

Figure~\ref{apext} illustrates comparisons between our CCD aperture 
magnitudes, and R-band magnitudes tabulated by other authors, for 
galaxies in common. The comparisons are quantified in 
Table~\ref{apextab}. 

\begin{table} 
\caption{ 
External comparison of aperture photometry with R-band work from 
other sources. Offsets are given in the sense `this work' -- 
`literature' .} \label{apextab} 
\begin{minipage}{80mm} 
\begin{tabular} {lrrr} 
\multicolumn{1}{l}{Source} & \multicolumn{1}{l}{$N_{gal}$} & \multicolumn{1}{l}{Mean offset} & \multicolumn{1}{l}{Dispersion} \\ 
\\ 
Steel (1997) & 22 & --0.002$\pm$0.010 & 0.047 \\ 
Postman \& Lauer (1995) & 5 & --0.025$\pm$0.011 & 0.026 \\ 
Colless et al. (1993) & 17 & --0.037$\pm$0.007 & 0.032 \\ 
\end{tabular} 
\end{minipage} 
\end{table} 

In the comparison with the photoelectric aperture photometry of 
Colless et al. (1993), we find a scatter which is well matched to the 
quadrature sum of our internal errors quoted above, and the similar 
uncertainties claimed by Colless et al. There exists, however, a small 
but significant offset of 0.037 mag. between the two datasets. 

Between our study and that of Postman and Lauer (1995, PL) there are 
six galaxies in common, of which one is severely contaminated by a 
star. We use only the remaining five galaxies to derive the quoted 
offset. (In performing this comparison, we have corrected for the 
misidentification by PL of the A0262 brightest cluster galaxy with 
N0705, rather than N0708.) 

Our data has been compared to the independent data of Steel (1997), 
which are also derived using the Harris R filter, but at a different 
telescope. The source of bimodality in this comparison has not been 
identified. 

In conclusion, small zero-point discrepancies do exist at the level of 
a few 0.01 mag. between the present photometric system and data from 
the literature. Internally, however, the data presented here possess 
a consistency of better than 0.01 mag.

\subsubsection{Derived parameters} 

\begin{figure*} 
\epsfig{file=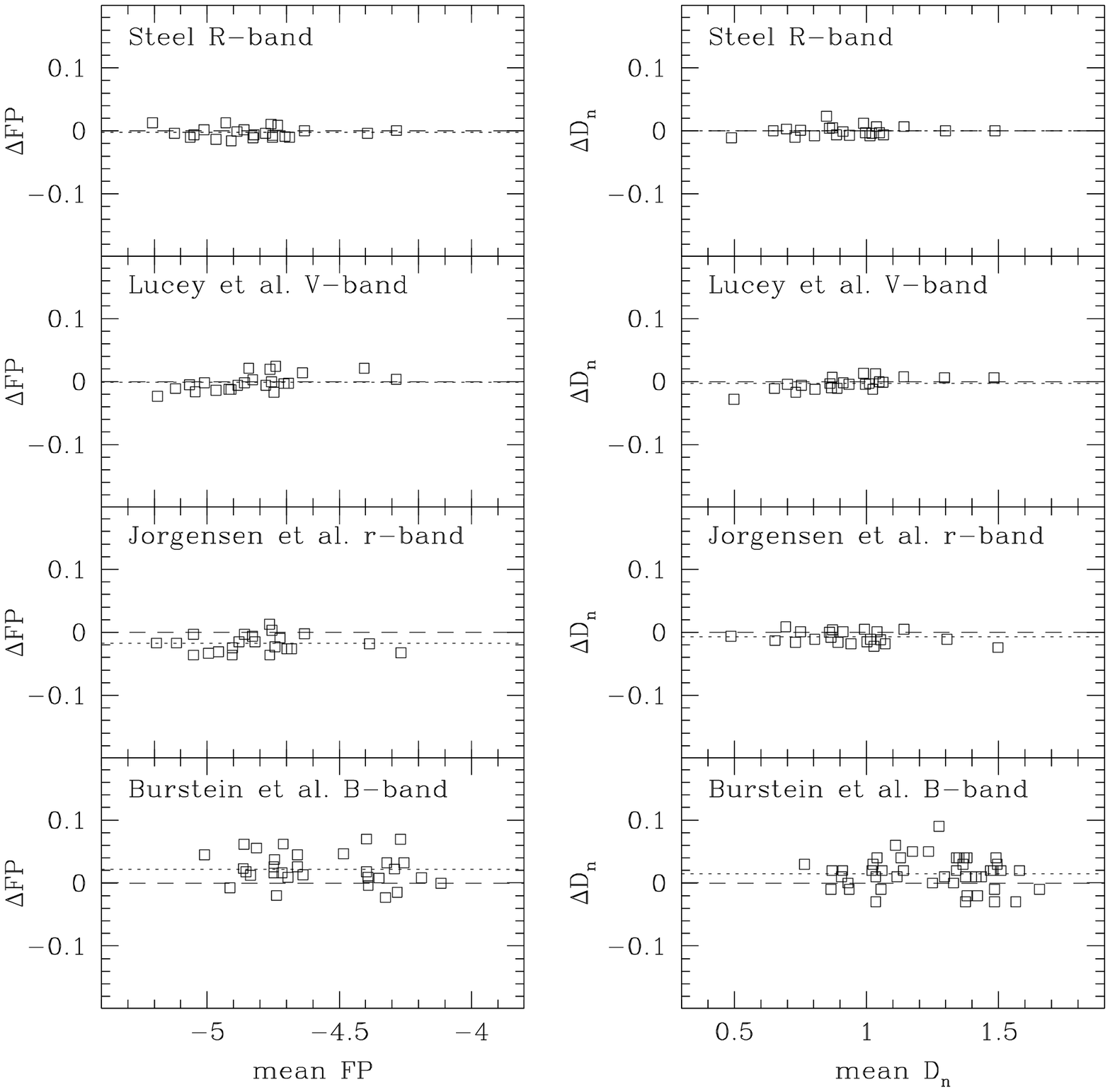, height=160mm, width=160mm} 
\caption{ 
External comparisons of \dn\ and FP combination ($\log A_{\rm e} - 
0.3\langle\mu_{\rm e}\rangle$). The comparison data are taken from 
Steel (1997) (R-band), Lucey et al. (1997) (V-band), J\o rgensen et 
al. (1995b) (r-band) and Burstein et al. (1987) (B-band). 
$\Delta\dn$ and $\Delta$FP are plotted in the sense `this work' -- 
`literature', against the mean of our measurement and the literature 
value. The dotted line indicates the mean offset in each panel.} 
\label{dnexfig} 
\end{figure*}

Figure~\ref{dnexfig} and Table~\ref{dnextab} present comparisons 
between the present data and photometric parameters from the 
literature, for galaxies in common. The comparison data are the 
R-band data of Steel (1997), the V-band measurements of Lucey et al. 
(1997), Gunn-r data from J\o rgensen et al. \shortcite{jfkphot}, and 
B-band parameters from Burstein et al. (1987). For the comparison with 
Burstein et al., only measurements given quality code `1' are 
included. The comparisons are performed for $\log\dn$ and for the 
FP combination, $\log A_{\rm e} - 0.3 \langle\mu\rangle_{\rm 
e}$. With the exception of the Burstein et al. data, the comparison 
galaxies are drawn entirely from the Coma cluster. The literature 
data are corrected for typical colours, according to the surface 
brightness level chosen by the author to define \dn. For instance, J\o 
rgensen et al. define an r-band \dn\ at $\langle\mu\rangle_{r}$ = 
19.60 mag. arcsec$^{-2}$. Surface brightnesses from their paper are 
therefore corrected for a colour of 0.37 mag. in $r - R$. 

The FP variables are compared in combination rather than separately, 
since the individual parameters $\log A_{\rm{e}}$ and 
$\langle\mu\rangle_{\rm e}$ can aquire correlated mean offsets from 
author to author, when the profile fit is performed over different 
ranges. The FP combination is, however, robust against changes to the 
range of fit.

\begin{table*} 
\caption{ 
Comparison of new R-band photometric parameters with measurements 
from Steel (1997), Lucey et al. (1997), J\o rgensen et al. (1995) 
and Burstein et al. (1987). For the Burstein et al. data, only 
measurements with quality code `1' are included in the comparison. 
Offsets are corrected for assumed mean colours, {\it viz.} $\langle\ 
V - R\ \rangle\ =\ 0.57$, $\langle\ r - R\ \rangle\ =\ 0.37$, 
$\langle\ 
B - R\ \rangle\ =\ 1.52$.} \label{dnextab} 
\begin{minipage}{130mm} 
\begin{tabular} {lrlrrr}

\multicolumn{1}{c}{Source} & \multicolumn{1}{c}{Parameter} & \multicolumn{1}{c}{$N_{gal}$} & \multicolumn{1}{c}{Mean Offset} & \multicolumn{1}{c}{Dispersion} \\ 
\\ 
Steel (1997) R-band & $\Delta (\log D_{n})$ & 22 & \phs\phm{+} 0.000 $\pm$ 0.001 & 0.007 \\ 
& $\Delta (\log A_{\rm e} - 0.3 \langle\mu\rangle_{\rm e})$ & 22 & \phm{+} --\ 0.002 $\pm$ 0.001 & 0.008 \\ 
\\ 
Lucey et al. (1997) V-band & $\Delta (\log D_{n})$ & 22 & \phm{+} --\ 0.003 $\pm$ 0.002 & 0.009 \\ 
& $\Delta (\log A_{\rm e} - 0.3 \langle\mu\rangle_{\rm e})$ & 22 & \phm{+} --\ 0.001 $\pm$ 0.002 & 0.013 \\ 
\\ 
J\o rgensen et al. (1995) r-band & $\Delta (\log D_{n})$ & 22 & \phm{+} --\ 0.007 $\pm$ 0.002 & 0.009 \\ 
& $\Delta (\log A_{\rm e} - 0.3 \langle\mu\rangle_{\rm e})$ & 22 & \phm{+} --\ 0.017 $\pm$ 0.002 & 0.014 \\ 
\\ 
Burstein et al. (1987) B-band & $\Delta (\log D_{n})$ & 47 & \phs +0.015 $\pm$ 0.003 & 0.025 \\ 
& $\Delta (\log A_{\rm e} - 0.3 \langle\mu\rangle_{\rm e})$ & 31 & \phs +0.022 $\pm$ 0.004 & 0.025 \\ 

\end{tabular} 
\end{minipage} 
\end{table*} 

The R-band photometry of Steel offers an independent validation of the 
present data, free from complications concerning band mis-matches, 
etc. The two samples agree to within 0.003 in both $\log\dn$ and the 
FP combination. 

From the excellent agreement of the R-band \dn\ measurements, 
presented here, with the V-band data of Lucey et al., we justify, {\it 
a posteriori}, the definition of our R-band \dn\ diameter at 
$\langle\mu\rangle_{R} = 19.23$ mag. The slight trend may be a 
reflection of the $V - R$ colour--magnitude relation for the Coma 
cluster. The slope found here (converted to magnitudes) is 
$-0.03\pm0.01$, which may be compared with the $V - K$ 
colour--magnitude slope of $-0.08\pm0.01$ reported by Bower et al. 
(1992). We note that the trend is in the expected sense, such that 
brighter galaxies are redder. 

The \dn\ comparison with the Gunn-r data of J\o rgensen et al. 
exhibits a curious bimodality. This is a result of their data being 
presented to only two decimal places in log\dn\, rather than three, as 
in our data. This is unfortunate, since their data is clearly more 
accurate than quoted, the dispersion given in Table~\ref{dnextab} 
being consequently overestimated.

The significant offset found between this work and that J\o rgensen et al. 
is, of course, sensitive to the adopted $r - R$ colour. 
From a comparison of our aperture 
magnitudes with magnitudes predicted from their tabulated r-band 
parameters, we derive a mean $r - R$ colour of 0.33 mag. If a colour 
correction were applied based on this result, the offsets of between 
this work and that of J\o rgensen et al. would be reduced to 
0.004$\pm$0.002 in $\log\dn$ and --0.006$\pm$0.003 in the FP 
combination. 

The photoelectric data of Burstein et al. have been corrected for the 
$(1+z)^{4}$ surface brightness dimming before comparison. The large 
offsets with respect to this source can be accounted for by the 
absence of a seeing correction in their data, as demonstrated by J\o 
rgensen et al.

The scatter in the comparisons is sufficiently small that our \dn\ / 
FP measurements may be brought onto a system consistent with external 
CCD data, to within 0.003 dex in implied distance. 

\bigskip 

\section{Conclusion} \label{conclusion} 
\mbox{} 

This paper has presented spectroscopic and photometric data to be used 
in a study of cluster peculiar motions in the Perseus--Pisces 
supercluster. The data comprise observations of 137 early-type 
galaxies in 9 clusters, and additional standard galaxies. 

From intermediate-dispersion spectroscopy, the velocity dispersion 
$\sigma$ has been derived for each galaxy, with a typical uncertainty 
of 7.6 per cent per measurement. The spectroscopic data also yield 
recession velocities $(cz)$ (to an uncertainty of about 30 \kms), and 
\mg\ indices (typical error 0.010 mag. per measurement). Extensive 
external comparisons are presented, allowing the $\sigma$ and \mg\ 
data to be placed onto a new `standard system', with an uncertainty of 
less than 0.01 dex.

R-band CCD photometry is used to derive global photometric parameters. 
The photometric data comprise effective diameter ($A_{\rm{e}}$), mean 
surface brightness within effective diameter 
($\langle\mu\rangle_{\rm{e}}$), and an R-band \dn\ parameter, defined 
analogously to the B-band photometric diameter of Dressler et al. 
(1987). The scatter of repeat observations indicates the following 
uncertainties -- $\log A_{\rm{e}}$: $\pm 0.032$ ; 
$\langle\mu\rangle_{\rm{e}}$: $\pm 0.113$ ; $\log\dn$: $\pm 0.005$; 
$\log A_{\rm{e}} - 0.3 \langle\mu\rangle_{\rm{e}}$: $\pm 0.006$. The 
aperture magnitudes, from which the profile is determined, show 
systematic offsets (at the level of a few 0.01 mag) with respect to 
literature data. The derived $\log\dn$ and Fundamental Plane parameter 
($\log A_{\rm{e}} - 0.3 \langle\mu\rangle_{\rm{e}}$), show a typical 
scatter of $\sim$0.010 with respect to similar data from the 
literature. 

The scatter in the FP relation is $\sim$0.08 dex, so that intrinsic 
scatter is dominant over random measurement errors. Currently, a major 
challenge in peculiar velocity work is to recognise and reduce the 
effects of {\it systematic} errors. We will defer until Paper II, a 
full discussion concerning such errors. For the present, we note that 
the high quality of the data presented here, together with the 
generous overlap secured with literature datasets, will allow us to 
address realistically the systematic errors in our peculiar velocity 
measurements. 

\section*{Acknowledgments} 

The Isaac Newton Telescope and Jacobus Kapteyn Telescope are operated 
on the island of La Palma by the Royal Greenwich Observatory in the 
Spanish Observatorio del Roque de los Muchachos of the Instituto de 
Astrof\'isica de Canarias. Data reduction was performed using 
Starlink facilities at Durham. JS and RJS acknowledge financial 
support from the PPARC. MJH acknowledges financial support from the 
PPARC; from a CITA National Fellowship; and from the Natural Sciences 
and Engineering Research Council of Canada, through operating 
grants to F. D. A. Hartwick and C. J. Pritchet. 
Alan Dressler is thanked for providing 
information for the subdivision of his Las Campanas datasets.

\clearpage 

\begin{table*} 
\begin{minipage}{105mm} 
\scriptsize 
\caption{ 
Raw spectroscopic data. In addition to our reference number for each 
galaxy, we tabulate under `Other ID' the relevant number from NGC, 
IC, UGC, CGCG catalogues, or from other published work. For each 
individual observation we list: the dataset from which values 
derive; $cz$ = heliocentric recession velocity; $\sigma$ = central 
velocity dispersion (kms$^{-1}$); 
$\varepsilon_{\sigma}$ = poisson error on 
$\sigma$; \mg\ = magnesium index (magnitudes) and $\varepsilon\sb{Mg_{2}}$ = 
Poisson error on \mg.} 
\label{rawspectable} 
\begin{tabular}{lllrrrrr} 
& & & & & & & \\ 
Our ID & Other ID & Dataset & cz & $\sigma$ & $\varepsilon_{\sigma}$ & \mg\ & $\varepsilon\sb{Mg_{2}}$ \\ 
& & & & & & & \\ 
{\bf Cluster : 7S21} \\ 
& & & & & & & \\ 
S01 & N0079 & TEK94 & 5479 & 194 & 11 & 0.307 & 0.012 \\ 
S02 & N0085A & TEK94 & 6189 & 108 & 6 & 0.239 & 0.012 \\ 
S03 & N0083 & TEK94 & 6263 & 253 & 14 & 0.321 & 0.013 \\ 
& & TEK94 & 6262 & 253 & 16 & - & - \\ 
S04 & N0080 & TEK94 & 5748 & 261 & 12 & 0.300 & 0.010 \\ 
& & TEK94 & 5734 & 249 & 13 & 0.305 & 0.009 \\ 
S05 & I1548 & TEK94 & 5775 & 149 & 6 & 0.197 & 0.008 \\ 
S06 & - & TEK94 & 5637 & 127 & 7 & 0.206 & 0.013 \\ 
& & TEK94 & 5655 & 131 & 13 & - & - \\ 
S07 & CGCG457-008 & TEK94 & 5926 & 115 & 8 & 0.254 & 0.012 \\ 
& & & & & & & \\ 
{\bf Cluster : Pisces} \\ 
& & & & & & & \\ 
Z01026 & N0398 & TEK94 & 4912 & 104 & 6 & 0.261 & 0.009 \\ 
Z01027 & N0379 & TEK94 & 5503 & 225 & 10 & 0.298 & 0.009 \\ 
& & EEV93 & 5492 & 243 & 16 & 0.287 & 0.012 \\ 
Z01030 & N0380 & EEV93 & 4448 & 314 & 12 & 0.329 & 0.007 \\ 
& & TEK94 & 4452 & 308 & 19 & 0.343 & 0.012 \\ 
& & EEV93 & 4441 & 337 & 15 & 0.328 & 0.009 \\ 
& & TEK94 & 4398 & 272 & 10 & 0.328 & 0.007 \\ 
& & TEK94 & 4442 & 301 & 11 & 0.343 & 0.007 \\ 
Z01032 & - & EEV93 & 4753 & 104 & 9 & 0.262 & 0.015 \\ 
Z01034 & CGCG501-077 & EEV93 & 5151 & 115 & 11 & 0.258 & 0.013 \\ 
& & TEK94 & 5152 & 127 & 5 & 0.267 & 0.008 \\ 
Z01035 & N0383 & EEV93 & 5082 & 269 & 11 & 0.293 & 0.008 \\ 
& & TEK94 & 5084 & 269 & 11 & 0.314 & 0.007 \\ 
& & EEV93 & 5009 & 249 & 22 & 0.299 & 0.014 \\ 
& & TEK94 & 5154 & 314 & 13 & 0.306 & 0.009 \\ 
& & TEK94 & 5117 & 316 & 18 & - & - \\ 
Z01035C1 & N0382 & EEV93 & 5243 & 222 & 10 & 0.275 & 0.009 \\ 
& & TEK94 & 5240 & 202 & 8 & 0.277 & 0.008 \\ 
& & EEV93 & 5233 & 195 & 16 & 0.246 & 0.015 \\ 
& & TEK94 & 5265 & 200 & 10 & 0.273 & 0.010 \\ 
Z01036 & I1618 & EEV93 & 4720 & 90 & 9 & 0.214 & 0.017 \\ 
Z01041 & N0386 & EEV93 & 5563 & 145 & 9 & 0.248 & 0.012 \\ 
& & TEK94 & 5548 & 121 & 7 & 0.247 & 0.011 \\ 
Z01043 & N0375 & EEV93 & 5910 & 180 & 10 & 0.265 & 0.010 \\ 
Z01045 & N0385 & EEV93 & 4988 & 198 & 8 & 0.271 & 0.009 \\ 
& & TEK94 & 5024 & 210 & 9 & 0.295 & 0.008 \\ 
& & TEK94 & 5027 & 195 & 10 & - & - \\ 
Z01046 & N0388 & TEK94 & 5473 & 148 & 8 & 0.251 & 0.009 \\ 
& & EEV93 & 5445 & 121 & 9 & 0.251 & 0.013 \\ 
Z01047 & - & TEK94 & 5493 & 132 & 6 & 0.285 & 0.009 \\ 
Z01049 & N0384 & EEV93 & 4255 & 257 & 10 & 0.309 & 0.009 \\ 
& & TEK94 & 4258 & 275 & 10 & 0.313 & 0.008 \\ 
& & EEV93 & 4267 & 245 & 12 & 0.291 & 0.010 \\ 
Z01073 & CGCG501-102 & EEV93 & 5174 & 172 & 10 & 0.276 & 0.011 \\ 
& & TEK94 & 5169 & 149 & 8 & 0.273 & 0.010 \\ 
Z02057 & N0420 & EEV93 & 5038 & 196 & 13 & 0.229 & 0.011 \\ 
& & TEK94 & 5005 & 179 & 7 & 0.250 & 0.007 \\ 
Z04035 & - & EEV93 & 23995 & 261 & 17 & 0.238 & 0.013 \\ 
Z04049 & N0394 & EEV93 & 4378 & 172 & 7 & 0.253 & 0.010 \\ 
& & TEK94 & 4404 & 195 & 8 & 0.260 & 0.008 \\ 
& & EEV93 & 4364 & 178 & 10 & 0.274 & 0.009 \\ 
Z04050 & N0392 & EEV93 & 4684 & 234 & 8 & 0.291 & 0.008 \\ 
& & TEK94 & 4665 & 231 & 7 & 0.292 & 0.006 \\ 
& & EEV93 & 4668 & 239 & 10 & 0.301 & 0.007 \\ 
Z04051 & N0397 & TEK94 & 4988 & 124 & 8 & 0.258 & 0.009 \\ 
Z05034 & I1638 & EEV93 & 4810 & 141 & 8 & 0.256 & 0.010 \\ 
& & TEK94 & 4814 & 162 & 7 & 0.276 & 0.008 \\ 
& & TEK94 & 4868 & 142 & 7 & 0.277 & 0.009 \\ 
& & TEK94 & 4811 & 137 & 7 & 0.278 & 0.010 \\ 
Z05044 & I1648 & TEK94 & 5541 & 124 & 8 & 0.266 & 0.010 \\ 
Z05052 & N0410 & EEV93 & 5315 & 292 & 11 & 0.344 & 0.007 \\ 
& & TEK94 & 5327 & 301 & 11 & 0.350 & 0.007 \\ 
& & TEK94 & 5309 & 310 & 12 & 0.346 & 0.007 \\ 
& & TEK94 & 5295 & 300 & 13 & - & - \\ 
Z10020 & - & TEK94 & 4852 & 85 & 7 & 0.227 & 0.013 \\ 
& & & & & & & \\ 
\end{tabular} 
\end{minipage} 
\end{table*} 

\begin{table*} 
\begin{minipage}{105mm} 
\scriptsize 
\contcaption{} 
\begin{tabular}{lllrrrrr} 
& & & & & & & \\ 
Our ID & Other ID & Dataset & cz & $\sigma$ & $\varepsilon_{\sigma}$ & \mg\ & $\varepsilon\sb{Mg_{2}}$ \\ 
& & & & & & & \\ 
Z14028 & CGCG501-070 & TEK94 & 4264 & 206 & 8 & 0.328 & 0.008 \\ 
& & EEV93 & 4252 & 192 & 10 & 0.307 & 0.009 \\ 
Z16012 & - & TEK94 & 17740 & 163 & 15 & 0.268 & 0.015 \\ 
Z17005 & - & TEK94 & 4651 & 105 & 6 & 0.205 & 0.010 \\ 
& & & & & & & \\ 
{\bf Cluster : HMS0122+3305} \\ 
& & & & & & & \\ 
H01022 & N0528 & EEV94 & 4806 & 245 & 9 & - & - \\ 
H01027 & - & TEK94 & 4976 & 99 & 9 & 0.210 & 0.011 \\ 
H01041 & N0499 & EEV94 & 4387 & 267 & 13 & - & - \\ 
& & TEK94 & 4399 & 259 & 10 & 0.327 & 0.006 \\ 
H01044 & N0501 & TEK94 & 5010 & 163 & 15 & 0.304 & 0.011 \\ 
H01051 & CGCG502-043 & EEV94 & 5225 & 153 & 9 & - & - \\ 
& & TEK94 & 5246 & 117 & 7 & 0.266 & 0.011 \\ 
H01056 & I1680 & TEK94 & 4438 & 136 & 6 & 0.267 & 0.010 \\ 
H01057 & N0508 & EEV94 & 5505 & 225 & 13 & - & - \\ 
& & TEK94 & 5526 & 228 & 12 & 0.310 & 0.009 \\ 
H01064 & N0507 & EEV94 & 4936 & 306 & 15 & - & - \\ 
& & TEK94 & 4937 & 290 & 11 & 0.299 & 0.006 \\ 
& & TEK94 & 4934 & 295 & 7 & 0.294 & 0.006 \\ 
H01075 & CGCG502-044 & TEK94 & 5141 & 133 & 8 & 0.278 & 0.009 \\ 
H01078 & I1673 & TEK94 & 5090 & 190 & 7 & 0.275 & 0.008 \\ 
H04010 & N0529 & EEV94 & 4815 & 236 & 9 & - & - \\ 
& & TEK94 & 4773 & 239 & 8 & 0.295 & 0.006 \\ 
& & & & & & & \\ 
{\bf Cluster : A0262} \\ 
& & & & & & & \\ 
A01043 & N0687 & EEV93 & 5112 & 204 & 10 & 0.276 & 0.011 \\ 
& & TEK94 & 5091 & 247 & 8 & 0.306 & 0.006 \\ 
A01047 & CGCG522-048 & TEK94 & 4151 & 144 & 7 & 0.263 & 0.008 \\ 
A01067 & N0703 & TEK94 & 5580 & 225 & 8 & 0.311 & 0.008 \\ 
A01069 & N0708 & TEK94 & 4855 & 219 & 16 & 0.321 & 0.016 \\ 
& & TEK94 & 4874 & 230 & 18 & 0.316 & 0.013 \\ 
A01071 & N0705 & EEV93 & 4514 & 183 & 12 & 0.258 & 0.011 \\ 
A01074 & N0704 & EEV93 & 4709 & 161 & 10 & 0.296 & 0.013 \\ 
& & TEK94 & 4730 & 159 & 6 & 0.275 & 0.007 \\ 
A01076 & - & EEV93 & 4284 & 125 & 13 & 0.275 & 0.020 \\ 
& & TEK94 & 4274 & 128 & 7 & 0.261 & 0.011 \\ 
A01094 & - & TEK94 & 14620 & 266 & 19 & 0.290 & 0.012 \\ 
A02025 & N0759 & EEV93 & 4667 & 259 & 14 & 0.247 & 0.011 \\ 
& & TEK94 & 4601 & 261 & 17 & 0.250 & 0.008 \\ 
A05096 & CGCG522-089 & TEK94 & 5245 & 92 & 10 & 0.220 & 0.014 \\ 
A05106 & N0732 & EEV93 & 5894 & 141 & 11 & 0.185 & 0.013 \\ 
A09029 & I0171 & EEV93 & 5360 & 173 & 11 & 0.227 & 0.012 \\ 
& & TEK94 & 5392 & 208 & 10 & 0.266 & 0.009 \\ 
A14050 & - & TEK94 & 5147 & 140 & 10 & 0.236 & 0.016 \\ 
A14078 & N0679 & TEK94 & 5049 & 254 & 9 & 0.305 & 0.007 \\ 
A19041 & U01269 & TEK94 & 3867 & 129 & 11 & 0.175 & 0.013 \\ 
& & TEK94 & 3829 & 105 & 9 & 0.195 & 0.012 \\ 
& & & & & & & \\ 
{\bf Cluster : A0347} \\ 
& & & & & & & \\ 
B02 & U01837 & TEK94 & 6582 & 197 & 13 & 0.302 & 0.012 \\ 
B03 & U01841 & TEK94 & 6373 & 235 & 15 & 0.304 & 0.011 \\ 
B03C & - & TEK94 & 6649 & 303 & 18 & 0.305 & 0.011 \\ 
B06 & U01859 & TEK94 & 5917 & 362 & 17 & 0.349 & 0.009 \\ 
B07 & CGCG538-065 & TEK94 & 5301 & 207 & 10 & 0.309 & 0.008 \\ 
B08 & N909 & TEK94 & 4978 & 191 & 12 & 0.271 & 0.009 \\ 
B09 & N910 & TEK94 & 5253 & 259 & 12 & 0.326 & 0.010 \\ 
& & TEK94 & 5222 & 243 & 13 & 0.342 & 0.010 \\ 
B10 & N911 & TEK94 & 5766 & 256 & 16 & 0.323 & 0.011 \\ 
B11 & N912 & TEK94 & 4418 & 175 & 9 & 0.290 & 0.010 \\ 
B16 & CGCG539-042 & TEK94 & 4885 & 156 & 7 & 0.267 & 0.010 \\ 
& & & & & & & \\ 
{\bf Cluster : J8} \\ 
& & & & & & & \\ 
J01049 & CGCG483-070 & EEV93 & 8555 & 316 & 24 & 0.298 & 0.012 \\ 
& & TEK94 & 8557 & 312 & 17 & 0.308 & 0.009 \\ 
& & EEV93 & 8556 & 299 & 34 & 0.298 & 0.015 \\ 
J01055 & - & TEK94 & 9613 & 143 & 13 & 0.283 & 0.020 \\ 
& & TEK94 & 9620 & 162 & 14 & 0.276 & 0.019 \\ 
J01056 & CGCG483-068 & EEV93 & 9438 & 212 & 17 & 0.333 & 0.016 \\ 
& & TEK94 & 9544 & 228 & 12 & 0.307 & 0.012 \\ 
& & TEK94 & 9549 & 253 & 13 & 0.316 & 0.011 \\ 
J01060 & I1803 & TEK94 & 9583 & 366 & 13 & 0.337 & 0.007 \\ 
J01065 & - & TEK94 & 9103 & 133 & 10 & 0.153 & 0.010 \\ 
J01067 & EFAR-J8-I & TEK94 & 9233 & 199 & 11 & 0.301 & 0.012 \\ 
& & EEV93 & 9232 & 197 & 21 & 0.290 & 0.017 \\ 
\end{tabular} 
\end{minipage} 
\end{table*} 

\begin{table*} 
\begin{minipage}{105mm} 
\scriptsize 
\contcaption{} 
\begin{tabular}{lllrrrrr} 
& & & & & & & \\ 
Our ID & Other ID & Dataset & cz & $\sigma$ & $\varepsilon_{\sigma}$ & \mg\ & $\varepsilon\sb{Mg_{2}}$ \\ 
& & & & & & & \\ 
J01069 & I1807 & TEK94 & 9094 & 199 & 9 & 0.266 & 0.009 \\ 
& & EEV93 & 9013 & 208 & 14 & 0.252 & 0.014 \\ 
J01070 & I1806 & EEV93 & 10190 & 177 & 23 & 0.306 & 0.017 \\ 
& & TEK94 & 10211 & 219 & 11 & 0.296 & 0.011 \\ 
& & EEV93 & 10236 & 245 & 24 & 0.286 & 0.018 \\ 
J01080 & - & TEK94 & 9731 & 164 & 12 & 0.240 & 0.012 \\ 
J03049 & - & EEV93 & 9929 & 264 & 27 & 0.244 & 0.017 \\ 
& & TEK94 & 9927 & 242 & 19 & 0.265 & 0.011 \\ 
J06039 & CGCG484-006 & TEK94 & 4268 & 135 & 5 & 0.273 & 0.007 \\ 
J07038 & - & TEK94 & 10136 & 182 & 11 & 0.269 & 0.012 \\ 
J08035 & - & TEK94 & 10099 & 81 & 11 & 0.167 & 0.017 \\ 
J08036 & EFAR-J8-K & EEV93 & 9770 & 170 & 16 & 0.278 & 0.018 \\ 
& & TEK94 & 9802 & 204 & 11 & 0.283 & 0.010 \\ 
& & EEV93 & 9817 & 196 & 14 & 0.267 & 0.018 \\ 
J09035 & - & TEK94 & 11133 & 282 & 17 & 0.325 & 0.010 \\ 

& & & & & & & \\ 
{\bf Cluster : Perseus (A0426)} \\ 
& & & & & & & \\ 
P01 & I0293 & EEV93 & 4704 & 150 & 11 & 0.260 & 0.015 \\ 
P02 & N1224 & TEK94 & 5235 & 247 & 10 & 0.270 & 0.009 \\ 
P03 & I0310 & TEK94 & 5660 & 218 & 12 & 0.249 & 0.010 \\ 
P05 & I0312 & EEV93 & 4978 & 222 & 13 & 0.296 & 0.012 \\ 
P07 & CR19 & TEK94 & 3544 & 123 & 10 & 0.239 & 0.015 \\ 
P08 & CR20 & TEK94 & 6454 & 188 & 13 & 0.271 & 0.017 \\ 
& & TEK94 & 6469 & 215 & 11 & 0.259 & 0.012 \\ 
P11 & BGP44 & TEK94 & 4247 & 159 & 14 & 0.275 & 0.014 \\ 
P12 & N1270 & EEV93 & 4965 & 351 & 16 & 0.350 & 0.010 \\ 
& & TEK94 & 5019 & 341 & 14 & 0.355 & 0.008 \\ 
P13 & PER195 & TEK94 & 8391 & 163 & 7 & 0.275 & 0.009 \\ 
& & EEV93 & 8392 & 193 & 20 & 0.283 & 0.018 \\ 
P14 & PER199 & EEV93 & 5078 & 226 & 16 & 0.275 & 0.016 \\ 
& & TEK94 & 5105 & 210 & 14 & 0.290 & 0.016 \\ 
& & TEK94 & 5113 & 213 & 7 & 0.279 & 0.008 \\ 
P15 & CR28 & TEK94 & 6213 & 212 & 7 & 0.288 & 0.009 \\ 
P16 & CR27 & EEV93 & 8053 & 171 & 13 & 0.266 & 0.012 \\ 
P17 & N1272 & TEK94 & 3802 & 272 & 16 & 0.331 & 0.011 \\ 
& & EEV93 & 3815 & 270 & 20 & 0.344 & 0.013 \\ 
& & EEV93 & 3794 & 286 & 20 & 0.340 & 0.013 \\ 
& & TEK94 & 3791 & 240 & 14 & 0.342 & 0.011 \\ 
& & EEV93 & 3801 & 278 & 15 & 0.334 & 0.014 \\ 
& & EEV94 & 3777 & 276 & 24 & - & - \\ 
& & TEK94 & 3794 & 263 & 11 & 0.338 & 0.011 \\ 
P18 & N1273 & EEV93 & 5387 & 207 & 15 & 0.249 & 0.013 \\ 
P19 & I1907 & EEV93 & 4479 & 195 & 18 & 0.278 & 0.016 \\ 
P20 & BGP111 & TEK94 & 3963 & 86 & 8 & 0.279 & 0.020 \\ 
P21 & PER152 & TEK94 & 3937 & 142 & 9 & 0.309 & 0.015 \\ 
& & TEK94 & 3943 & 136 & 9 & 0.315 & 0.014 \\ 
P22 & CR36 & EEV93 & 7460 & 202 & 14 & 0.280 & 0.012 \\ 
P23 & N1278 & EEV93 & 6044 & 238 & 15 & 0.292 & 0.011 \\ 
& & EEV94 & 6074 & 277 & 19 & - & - \\ 
& & TEK94 & 6061 & 270 & 14 & 0.306 & 0.011 \\ 
P26 & BGP59 & TEK94 & 5315 & 207 & 9 & 0.283 & 0.010 \\ 
P27 & U02673 & EEV93 & 4424 & 197 & 14 & 0.288 & 0.013 \\ 
P28 & N1281 & TEK94 & 4300 & 276 & 12 & 0.324 & 0.010 \\ 
P29 & N1282 & TEK94 & 2210 & 213 & 7 & 0.292 & 0.009 \\ 
& & EEV93 & 2223 & 203 & 11 & 0.290 & 0.012 \\ 
& & EEV93 & 2226 & 210 & 14 & 0.264 & 0.012 \\ 
& & TEK94 & 2216 & 228 & 9 & 0.277 & 0.009 \\ 
P30 & N1283 & EEV93 & 6744 & 224 & 12 & 0.277 & 0.012 \\ 
& & EEV94 & 6735 & 204 & 13 & - & - \\ 
P31 & PER163 & TEK94 & 5483 & 200 & 11 & 0.293 & 0.013 \\ 
& & TEK94 & 5480 & 164 & 7 & 0.278 & 0.008 \\ 
P33 & BGP33 & TEK94 & 4950 & 168 & 8 & 0.289 & 0.009 \\ 
P34 & I0313 & TEK94 & 4432 & 242 & 11 & 0.331 & 0.009 \\ 
P36 & N1293 & EEV93 & 4170 & 216 & 12 & 0.293 & 0.011 \\ 
& & EEV93 & 4149 & 218 & 23 & 0.307 & 0.015 \\ 
& & TEK94 & 4164 & 208 & 8 & 0.319 & 0.008 \\ 
P37 & U02698 & EEV93 & 6472 & 373 & 22 & 0.318 & 0.012 \\ 
& & TEK94 & 6421 & 364 & 14 & 0.340 & 0.009 \\ 
P38 & U02717 & EEV93 & 3778 & 165 & 13 & 0.239 & 0.015 \\ 
& & TEK94 & 3798 & 153 & 9 & 0.227 & 0.011 \\ 
P39 & U02725 & TEK94 & 6215 & 220 & 10 & 0.293 & 0.008 \\ 
\end{tabular} 
\end{minipage} 
\end{table*} 

\begin{table*} 
\begin{minipage}{105mm} 
\scriptsize 
\contcaption{} 
\begin{tabular}{lllrrrrr} 
& & & & & & & \\ 
Our ID & Other ID & Dataset & cz & $\sigma$ & $\varepsilon_{\sigma}$ & \mg\ & $\varepsilon\sb{Mg_{2}}$ \\ 
& & & & & & & \\ 
{\bf Cluster : Coma (A1656)} \\ 
& & & & & & & \\ 
N4875 & COMA-D104 & EEV93 & 8047 & 168 & 13 & 0.272 & 0.014 \\ 
N4886 & COMA-D151 & EEV93 & 6372 & 167 & 8 & 0.252 & 0.014 \\ 
N4860 & COMA-D194 & EEV93 & 7944 & 312 & 27 & 0.324 & 0.015 \\ 
N4881 & COMA-D217 & EEV93 & 6732 & 166 & 14 & 0.270 & 0.016 \\ 
I4011 & COMA-D150 & EEV93 & 7233 & 118 & 13 & 0.260 & 0.024 \\ 
COMA-D125 & - & EEV93 & 6910 & 174 & 19 & 0.232 & 0.017 \\ 
& & & & & & & \\ 
{\bf Cluster : A2199} \\ 
& & & & & & & \\ 
A21-F113 & - & TEK94 & 7995 & 169 & 14 & 0.242 & 0.016 \\ 
& & TEK94 & 8068 & 163 & 12 & 0.263 & 0.013 \\ 
& & TEK94 & 8017 & 170 & 9 & 0.254 & 0.012 \\ 
& & TEK94 & 8128 & 158 & 15 & 0.262 & 0.013 \\ 
A21-F114 & - & TEK94 & 9177 & 199 & 12 & 0.294 & 0.013 \\ 
& & TEK94 & 9179 & 212 & 12 & 0.288 & 0.011 \\ 
& & TEK94 & 9128 & 191 & 10 & 0.278 & 0.010 \\ 
& & TEK94 & 9255 & 192 & 9 & 0.300 & 0.010 \\ 
A21-F121 & A2199-S26 & TEK94 & 8783 & 177 & 15 & 0.270 & 0.013 \\ 
& & TEK94 & 8754 & 170 & 11 & 0.278 & 0.012 \\ 
A21-F144 & A2199-S30 & TEK94 & 8510 & 256 & 12 & 0.244 & 0.010 \\ 
& & TEK94 & 8539 & 259 & 9 & 0.254 & 0.008 \\ 
A21-F145 & - & TEK94 & 7586 & 152 & 8 & 0.270 & 0.012 \\ 
& & TEK94 & 7634 & 146 & 8 & 0.282 & 0.010 \\ 
A21-F146 & A2199-S34 & TEK94 & 8302 & 158 & 9 & 0.263 & 0.014 \\ 
& & TEK94 & 8327 & 163 & 11 & 0.254 & 0.011 \\ 
A21-F164 & N6166 & TEK94 & 9329 & 269 & 25 & 0.321 & 0.014 \\ 
& & TEK94 & 9367 & 285 & 18 & 0.298 & 0.011 \\ 
A21-Z34A & A2199-Z34A & TEK94 & 8724 & 208 & 10 & 0.260 & 0.008 \\ 
& & TEK94 & 8718 & 190 & 9 & 0.289 & 0.009 \\ 
A21-Z34AC & - & TEK94 & 8949 & 227 & 9 & 0.297 & 0.008 \\ 
& & TEK94 & 8980 & 208 & 8 & 0.325 & 0.009 \\ 
N6158 & - & TEK94 & 8936 & 197 & 14 & 0.272 & 0.015 \\ 
& & TEK94 & 8925 & 186 & 9 & 0.264 & 0.011 \\ 
& & TEK94 & 8963 & 174 & 7 & 0.280 & 0.008 \\ 
& & & & & & & \\ 
{\bf Cluster : A2634} \\ 
& & & & & & & \\ 
A26-F102 & A2634-D107 & TEK94 & 9298 & 213 & 13 & - & - \\ 
A26-F1201 & A2634-D79 & TEK94 & 10156 & 188 & 12 & 0.272 & 0.014 \\ 
& & EEV93 & 10166 & 118 & 10 & 0.298 & 0.017 \\ 
& & EEV93 & 10178 & 146 & 21 & 0.270 & 0.023 \\ 
& & TEK94 & 10144 & 169 & 11 & 0.275 & 0.013 \\ 
A26-F121 & A2634-D80 & TEK94 & 9547 & 206 & 16 & 0.284 & 0.014 \\ 
& & EEV93 & 9571 & 157 & 13 & 0.250 & 0.016 \\ 
& & EEV93 & 9595 & 163 & 20 & 0.257 & 0.023 \\ 
& & TEK94 & 9595 & 186 & 12 & 0.270 & 0.013 \\ 
A26-F1221 & N7720 & EEV93 & 9117 & 354 & 23 & 0.308 & 0.010 \\ 
& & TEK94 & 9110 & 322 & 16 & 0.322 & 0.010 \\ 
& & TEK94 & 9105 & 340 & 16 & 0.310 & 0.009 \\ 
& & EEV93 & 9091 & 310 & 16 & 0.305 & 0.009 \\ 
& & TEK94 & 9079 & 349 & 15 & 0.320 & 0.007 \\ 
A26-F1222 & A2634-D76 & EEV93 & 8107 & 230 & 15 & 0.276 & 0.013 \\ 
& & TEK94 & 8091 & 209 & 13 & 0.278 & 0.013 \\ 
& & TEK94 & 8143 & 176 & 14 & 0.289 & 0.012 \\ 
& & EEV93 & 8139 & 209 & 11 & 0.274 & 0.013 \\ 
& & TEK94 & 8115 & 230 & 15 & 0.275 & 0.009 \\ 
A26-F129 & A2634-D74 & EEV93 & 8423 & 199 & 16 & 0.267 & 0.015 \\ 
& & TEK94 & 8420 & 213 & 11 & 0.289 & 0.010 \\ 
& & TEK94 & 8422 & 202 & 10 & 0.290 & 0.011 \\ 
A26-F134 & A2634-D55 & EEV93 & 9281 & 221 & 14 & 0.279 & 0.012 \\ 
& & TEK94 & 9287 & 232 & 10 & 0.284 & 0.008 \\ 
& & TEK94 & 9286 & 206 & 11 & 0.287 & 0.009 \\ 
A26-F138 & A2634-D58 & EEV93 & 10883 & 240 & 18 & 0.287 & 0.012 \\ 
& & TEK94 & 10849 & 221 & 13 & 0.295 & 0.012 \\ 
& & TEK94 & 10924 & 220 & 12 & 0.316 & 0.011 \\ 
& & EEV93 & 10815 & 195 & 22 & 0.287 & 0.022 \\ 
A26-F139 & A2634-D57 & EEV93 & 9604 & 206 & 11 & 0.300 & 0.012 \\ 
& & TEK94 & 9563 & 212 & 13 & 0.309 & 0.012 \\ 
& & TEK94 & 9633 & 216 & 11 & 0.335 & 0.011 \\ 
& & EEV93 & 9545 & 217 & 26 & 0.277 & 0.021 \\ 
A26-F1482 & A2634-D38 & TEK94 & 9345 & 240 & 22 & - & - \\ 
\end{tabular} 
\end{minipage} 
\end{table*} 

\begin{table*} 
\begin{minipage}{105mm} 
\scriptsize 
\contcaption{} 
\begin{tabular}{lllrrrrr} 
& & & & & & & \\ 
Our ID & Other ID & Dataset & cz & $\sigma$ & $\varepsilon_{\sigma}$ & \mg\ & $\varepsilon\sb{Mg_{2}}$ \\ 
& & & & & & & \\ 
{\bf Field and standards} \\ 
& & & & & & & \\ 
N0541 & - & TEK94 & 5443 & 218 & 13 & 0.307 & 0.010 \\ 
N0545 & - & TEK94 & 5341 & 244 & 10 & 0.303 & 0.008 \\ 
N0547 & - & TEK94 & 5545 & 250 & 12 & 0.314 & 0.008 \\ 
N0548 & - & TEK94 & 5410 & 148 & 8 & 0.237 & 0.011 \\ 
N0584 & - & TEK94 & 1833 & 205 & 11 & 0.291 & 0.009 \\ 
& & TEK94 & 1830 & 196 & 7 & 0.267 & 0.007 \\ 
N0596 & - & TEK94 & 1872 & 162 & 5 & 0.251 & 0.006 \\ 
N0621 & U01147 & EEV93 & 5086 & 198 & 12 & 0.273 & 0.013 \\ 
N0661 & U01215 & EEV93 & 3827 & 197 & 7 & 0.288 & 0.008 \\ 
& & EEV94 & 3817 & 186 & 6 & - & - \\ 
N0680 & U01286 & EEV93 & 2963 & 210 & 8 & 0.279 & 0.008 \\ 
& & EEV94 & 2917 & 196 & 8 & - & - \\ 
N0741 & - & TEK94 & 5545 & 264 & 23 & 0.343 & 0.029 \\ 
N0770 & U01463 & EEV93 & 2569 & 111 & 10 & 0.213 & 0.012 \\ 
& & EEV94 & 2505 & 126 & 9 & - & - \\ 
N0821 & - & TEK94 & 1753 & 208 & 6 & 0.316 & 0.009 \\ 
& & TEK94 & 1758 & 196 & 13 & 0.304 & 0.014 \\ 
N0936 & - & TEK94 & 1439 & 205 & 8 & 0.302 & 0.008 \\ 
& & EEV94 & 1306 & 182 & 9 & - & - \\ 
& & TEK94 & 1331 & 183 & 12 & 0.300 & 0.014 \\ 
N0968 & U02040 & EEV93 & 3627 & 217 & 11 & 0.267 & 0.011 \\ 
& & EEV94 & 3599 & 201 & 9 & - & - \\ 
N1023 & - & EEV93 & 614 & 194 & 18 & 0.326 & 0.014 \\ 
& & EEV93 & 622 & 205 & 10 & 0.324 & 0.009 \\ 
N1198 & U02533 & EEV93 & 1592 & 74 & 6 & 0.113 & 0.008 \\ 

N3377 & - & EEV93 & 655 & 174 & 7 & 0.274 & 0.006 \\ 
& & EEV93 & 679 & 129 & 7 & 0.266 & 0.007 \\ 
N3379 & - & EEV93 & 896 & 216 & 5 & 0.313 & 0.004 \\ 
N3384 & - & EEV93 & 735 & 171 & 4 & 0.310 & 0.006 \\ 
N3412 & - & EEV93 & 857 & 104 & 4 & 0.238 & 0.007 \\ 
N3489 & - & EEV93 & 690 & 96 & 4 & 0.188 & 0.005 \\ 
N3862 & - & EEV93 & 6498 & 258 & 15 & 0.297 & 0.013 \\ 
N4472 & - & EEV93 & 966 & 270 & 18 & 0.307 & 0.017 \\ 
N4478 & - & EEV93 & 1356 & 159 & 11 & 0.270 & 0.022 \\ 
N4564 & - & EEV93 & 1158 & 191 & 19 & 0.380 & 0.015 \\ 
N6173 & - & TEK94 & 8790 & 292 & 13 & 0.295 & 0.007 \\ 
& & TEK94 & 8870 & 279 & 16 & 0.287 & 0.010 \\ 
N6411 & - & TEK94 & 3756 & 192 & 13 & 0.277 & 0.011 \\ 
& & TEK94 & 3845 & 175 & 8 & 0.269 & 0.009 \\ 
N6482 & - & TEK94 & 3841 & 304 & 12 & 0.333 & 0.009 \\ 
& & TEK94 & 3931 & 295 & 14 & 0.329 & 0.016 \\ 
N6702 & - & TEK94 & 4761 & 196 & 12 & 0.263 & 0.013 \\ 
& & TEK94 & 4748 & 169 & 9 & 0.262 & 0.011 \\ 
& & TEK94 & 4739 & 177 & 16 & 0.293 & 0.012 \\ 
N6703 & - & TEK94 & 2393 & 190 & 12 & 0.281 & 0.010 \\ 
& & TEK94 & 2408 & 201 & 8 & 0.265 & 0.007 \\ 
& & TEK94 & 2388 & 195 & 7 & 0.281 & 0.007 \\ 
N7236 & - & TEK94 & 7879 & 247 & 10 & 0.270 & 0.008 \\ 
N7237 & - & TEK94 & 7868 & 203 & 11 & 0.312 & 0.014 \\ 
N7385 & - & TEK94 & 7856 & 282 & 15 & 0.321 & 0.009 \\ 
N7391 & - & TEK94 & 3048 & 259 & 9 & 0.324 & 0.009 \\ 
N7454 & - & TEK94 & 2058 & 105 & 6 & 0.202 & 0.010 \\ 
& & TEK94 & 1988 & 99 & 8 & - & - \\ 
N7562 & - & TEK94 & 3657 & 250 & 12 & 0.277 & 0.009 \\ 
N7617 & - & TEK94 & 4172 & 130 & 8 & 0.217 & 0.013 \\ 
N7619 & - & TEK94 & 3814 & 337 & 12 & 0.335 & 0.007 \\ 
& & EEV94 & 3836 & 338 & 14 & - & - \\ 
N7626 & - & TEK94 & 3425 & 281 & 13 & 0.322 & 0.010 \\ 
& & EEV94 & 3454 & 274 & 7 & - & - \\ 
N7768 & - & TEK94 & 8179 & 339 & 14 & 0.299 & 0.008 \\ 
I2955 & - & EEV93 & 6478 & 245 & 20 & 0.228 & 0.019 \\ 
U02554 & - & TEK94 & 2863 & 135 & 19 & 0.261 & 0.017 \\ 
U03115 & - & TEK94 & 3255 & 120 & 18 & 0.119 & 0.021 \\ 
Q05 & CGCG477-023 & TEK94 & 8432 & 204 & 11 & 0.296 & 0.010 \\ 
& & & & & & & \\ 
\end{tabular} 
\end{minipage} 
\end{table*}

\begin{table*} 
\begin{minipage}{110mm} 
\scriptsize 
\caption{ 
Photometric data. Together with identification numbers, we tabulate: 
$R_{20}$ = raw magnitude within 20 arcsec diameter aperture; $A_{B}$ 
= B-band galactic extinction; psf = FWHM seeing (arcsec), as 
measured from 
stellar images; $\log A_{\rm{e}}$ = log effective diameter (arcsec); 
$\langle\mu\rangle_{\rm{e}}$ = mean surface brightness (mag. arcsec$^{-2}$) 
within $A_{\rm{e}}$; rms = rms residual of galaxy profile to best-fit 
$R^{1/4}$ law (magnitudes); log\dn\ = log R-band photometric \dn\ parameter 
(arcsec).} 
\label{rawphotable} 
\begin{tabular}{llrrrrrrr} 
& & & & & & & \\ 
Our ID & Other ID & $R_{20}$ & $A_{B}$ & psf & $\log A_{\rm{e}}$ & $\langle\mu\rangle_{\rm{e}}$ & rms & $\log D\sb{n}$ \\ 
& & & & & & & \\ 
{\bf Cluster : 7S21} \\ 
& & & & & & & \\ 
S01 & N0079 & 13.69 & 0.05 & 1.8 & 1.354 & 20.00 & 0.01 & 1.132 \\ 
S02 & N0085A & 14.16 & 0.05 & 1.5 & 1.502 & 20.99 & 0.03 & 0.953 \\ 
S03 & N0083 & 13.10 & 0.09 & 1.5 & 1.659 & 20.43 & 0.04 & 1.318 \\ 
S04 & N0080 & 12.95 & 0.09 & 1.6 & 1.691 & 20.28 & 0.04 & 1.364 \\ 
S05 & I1548 & 14.04 & 0.09 & 1.2 & 0.992 & 18.91 & 0.03 & 1.072 \\ 
S06 & - & 14.88 & 0.09 & 1.0 & 1.128 & 20.34 & 0.01 & 0.796 \\ 
S07 & CGCG457-008 & 14.07 & 0.11 & 1.3 & 1.128 & 19.47 & 0.02 & 1.055 \\ 
& & & & & & & \\ 
{\bf Cluster : Pisces} \\ 
& & & & & & & \\ 
Z01026 & N0398 & 14.10 & 0.18 & 1.1 & 1.155 & 19.63 & 0.01 & 1.046 \\ 
Z01027 & N0379 & 12.92 & 0.17 & 1.2 & 1.518 & 19.81 & 0.07 & 1.380 \\ 
& & 12.94 & 0.17 & 1.5 & 1.509 & 19.79 & 0.06 & 1.375 \\ 
& & 12.94 & 0.17 & 1.2 & 1.600 & 20.06 & 0.05 & 1.374 \\ 
Z01030 & N0380 & 12.86 & 0.17 & 1.1 & 1.341 & 19.03 & 0.03 & 1.385 \\ 
& & 12.85 & 0.17 & 1.5 & 1.307 & 18.90 & 0.02 & 1.387 \\ 
& & 12.85 & 0.17 & 1.2 & 1.309 & 18.90 & 0.02 & 1.388 \\ 
Z01032 & - & 14.78 & 0.18 & 2.0 & 1.100 & 20.08 & 0.01 & 0.852 \\ 
Z01034 & CGCG501-077 & 14.19 & 0.16 & 1.0 & 1.192 & 19.79 & 0.04 & 1.026 \\ 
Z01035 & N0383 & 12.59 & 0.17 & 1.0 & 1.788 & 20.27 & 0.04 & 1.483 \\ 
& & 12.58 & 0.17 & 1.2 & 1.791 & 20.27 & 0.04 & 1.486 \\ 
Z01035C1 & N0382 & 13.64 & 0.17 & 1.0 & 1.093 & 18.93 & 0.04 & 1.186 \\ 
& & 13.63 & 0.17 & 1.2 & 1.079 & 18.87 & 0.04 & 1.190 \\ 
Z01036 & I1618 & 14.32 & 0.16 & 1.2 & 1.242 & 20.16 & 0.02 & 0.967 \\ 
& & 14.34 & 0.16 & 2.7 & 1.252 & 20.21 & 0.02 & 0.963 \\ 
Z01041 & N0386 & 14.11 & 0.17 & 1.0 & 1.042 & 19.25 & 0.06 & 1.044 \\ 
& & 14.16 & 0.17 & 0.8 & 1.107 & 19.54 & 0.05 & 1.028 \\ 
Z01043 & N0375 & 14.27 & 0.17 & 0.8 & 0.840 & 18.50 & 0.01 & 1.035 \\ 
Z01045 & N0385 & 13.33 & 0.17 & 1.1 & 1.408 & 19.72 & 0.02 & 1.258 \\ 
& & 13.31 & 0.17 & 0.7 & 1.409 & 19.70 & 0.02 & 1.263 \\ 
& & 13.35 & 0.17 & 1.4 & 1.380 & 19.65 & 0.02 & 1.253 \\ 
& & 13.34 & 0.17 & 1.4 & 1.386 & 19.66 & 0.02 & 1.254 \\ 
Z01046 & N0388 & 14.22 & 0.18 & 0.8 & 0.905 & 18.72 & 0.02 & 1.049 \\ 
& & 14.28 & 0.18 & 1.7 & 0.857 & 18.61 & 0.03 & 1.030 \\ 
Z01047 & - & 14.44 & 0.17 & 0.8 & 0.839 & 18.66 & 0.01 & 0.991 \\ 
Z01049 & N0384 & 13.24 & 0.17 & 1.1 & 1.187 & 18.88 & 0.02 & 1.281 \\ 
& & 13.24 & 0.17 & 1.4 & 1.171 & 18.82 & 0.01 & 1.281 \\ 
& & 13.23 & 0.17 & 1.4 & 1.212 & 18.95 & 0.02 & 1.283 \\ 
Z01073 & CGCG501-102 & 14.08 & 0.18 & 1.1 & 1.072 & 19.22 & 0.04 & 1.060 \\ 
& & 14.07 & 0.18 & 1.2 & 1.050 & 19.11 & 0.04 & 1.064 \\ 
Z02057 & N0420 & 13.08 & 0.16 & 1.1 & 1.523 & 19.87 & 0.02 & 1.330 \\ 
& & 13.06 & 0.16 & 1.5 & 1.542 & 19.90 & 0.03 & 1.337 \\ 
& & 13.06 & 0.16 & 1.6 & 1.539 & 19.90 & 0.02 & 1.337 \\ 
& & 13.05 & 0.16 & 1.2 & 1.521 & 19.83 & 0.02 & 1.339 \\ 
Z04049 & N0394 & 13.61 & 0.18 & 1.1 & 1.074 & 18.82 & 0.01 & 1.188 \\ 
& & 13.64 & 0.18 & 1.8 & 1.011 & 18.62 & 0.03 & 1.183 \\ 
& & 13.63 & 0.18 & 1.4 & 1.070 & 18.83 & 0.01 & 1.181 \\ 
Z04050 & N0392 & 13.00 & 0.18 & 1.1 & 1.366 & 19.26 & 0.02 & 1.351 \\ 
& & 13.01 & 0.18 & 1.8 & 1.382 & 19.32 & 0.02 & 1.351 \\ 
& & 13.01 & 0.18 & 1.4 & 1.373 & 19.29 & 0.02 & 1.350 \\ 
Z04051 & N0397 & 14.36 & 0.18 & 1.0 & 1.005 & 19.25 & 0.01 & 0.997 \\ 
& & 14.34 & 0.18 & 1.8 & 1.032 & 19.32 & 0.02 & 1.003 \\ 
& & 14.35 & 0.18 & 1.4 & 1.011 & 19.26 & 0.02 & 1.000 \\ 
Z05034 & I1638 & 13.80 & 0.16 & 1.1 & 1.236 & 19.60 & 0.03 & 1.119 \\ 
& & 13.80 & 0.16 & 1.4 & 1.252 & 19.66 & 0.03 & 1.119 \\ 
Z05044 & I1648 & 14.07 & 0.17 & 1.7 & 1.116 & 19.47 & 0.03 & 1.050 \\ 
Z05052 & N0410 & 12.50 & 0.19 & 0.8 & 1.732 & 19.93 & 0.02 & 1.524 \\ 
& & 12.50 & 0.19 & 1.7 & 1.737 & 19.93 & 0.03 & 1.524 \\ 
Z10020 & CGCG501-126 & 14.62 & 0.20 & 1.4 & 1.134 & 20.08 & 0.03 & 0.879 \\ 
Z14028 & CGCG501-070 & 13.78 & 0.18 & 1.2 & 0.928 & 18.39 & 0.02 & 1.143 \\ 
& & 13.80 & 0.18 & 1.6 & 0.928 & 18.40 & 0.01 & 1.142 \\ 
Z16012 & - & 15.36 & 0.14 & 1.3 & 0.941 & 19.80 & 0.01 & 0.779 \\ 
Z17005 & - & 14.47 & 0.17 & 1.5 & 0.937 & 19.14 & 0.03 & 0.969 \\ 
& & & & & & & \\ 
{\bf Cluster : HMS0122+3305} \\ 
& & & & & & & \\ 
H01022 & N0528 & 13.05 & 0.17 & 1.2 & 1.330 & 19.20 & 0.01 & 1.338 \\ 
H01041 & N0499 & 12.55 & 0.17 & 1.2 & 1.545 & 19.44 & 0.01 & 1.481 \\ 
H01044 & N0501 & 14.14 & 0.16 & 1.2 & 1.029 & 19.16 & 0.02 & 1.041 \\ 
H01051 & CGCG502-043 & 14.08 & 0.13 & 1.1 & 1.100 & 19.41 & 0.01 & 1.048 \\ 
H01056 & I1680 & 14.03 & 0.17 & 1.1 & 1.035 & 19.05 & 0.03 & 1.073 \\ 
\end{tabular} 
\end{minipage} 
\end{table*} 

\begin{table*} 
\begin{minipage}{100mm} 
\scriptsize 
\contcaption{} 
\begin{tabular}{llrrrrrrr} 
& & & & & & & \\ 
Our ID & Other ID & $R_{20}$ & $A_{B}$ & psf & $\log A_{\rm{e}}$ & $\langle\mu\rangle_{\rm{e}}$ & rms & $\log D\sb{n}$ \\ 
H01057 & N0508 & 13.43 & 0.16 & 1.0 & 1.445 & 19.96 & 0.02 & 1.224 \\ 
H01064 & N0507 & 12.48 & 0.16 & 1.0 & 1.668 & 19.70 & 0.05 & 1.512 \\ 
H01078 & I1673 & 13.81 & 0.13 & 1.0 & 0.856 & 18.14 & 0.01 & 1.139 \\ 
H04010 & N0529 & 12.64 & 0.15 & 1.1 & 1.424 & 19.14 & 0.01 & 1.448 \\ 
& & & & & & & \\ 
{\bf Cluster : A0262} \\ 
& & & & & & & \\ 
A01043 & N0687 & 12.79 & 0.21 & 1.2 & 1.478 & 19.41 & 0.02 & 1.420 \\ 
A01047 & CGCG522-048 & 13.86 & 0.24 & 1.1 & 1.327 & 19.95 & 0.03 & 1.104 \\ 
A01067 & N0703 & 13.47 & 0.24 & 1.8 & 1.361 & 19.69 & 0.01 & 1.233 \\ 
A01069 & N0708 & 13.27 & 0.24 & 1.8 & 2.095 & 21.75 & 0.05 & 1.282 \\ 
A01071 & N0705 & 13.57 & 0.24 & 1.8 & 1.181 & 19.19 & 0.04 & 1.209 \\ 
A01074 & N0704 & 13.76 & 0.24 & 1.7 & 1.153 & 19.22 & 0.01 & 1.155 \\ 
A01076 & - & 14.55 & 0.24 & 1.7 & 1.154 & 20.02 & 0.02 & 0.931 \\ 
A01094 & - & 14.66 & 0.18 & 1.1 & 1.030 & 19.48 & 0.02 & 0.955 \\ 
A02025 & N0759 & 12.99 & 0.20 & 1.1 & 1.543 & 19.85 & 0.01 & 1.363 \\ 
A05096 & CGCG522-089 & 14.59 & 0.18 & 1.1 & 1.290 & 20.55 & 0.03 & 0.897 \\ 
A09029 & I0171 & 12.90 & 0.20 & 1.4 & 1.777 & 20.47 & 0.02 & 1.407 \\ 
A19041 & U01269 & 14.32 & 0.13 & 1.3 & 1.479 & 21.05 & 0.01 & 0.910 \\ 
& & & & & & & \\ 
{\bf Cluster : A0347} \\ 
& & & & & & & \\ 
B02 & U01837 & 13.52 & 0.34 & 1.4 & 1.589 & 20.43 & 0.01 & 1.234 \\ 
B03 & U01841 & 13.18 & 0.34 & 1.4 & 1.826 & 20.77 & 0.02 & 1.344 \\ 
& & 13.22 & 0.34 & 1.4 & 1.799 & 20.74 & 0.02 & 1.329 \\ 
B03C & - & 14.51 & 0.34 & 1.4 & 0.575 & 17.45 & 0.01 & 1.030 \\ 
& & 14.52 & 0.34 & 1.4 & 0.601 & 17.57 & 0.00 & 1.024 \\ 
B06 & U01859 & 13.20 & 0.31 & 1.4 & 1.182 & 18.68 & 0.02 & 1.321 \\ 
B07 & CGCG538-065 & 13.66 & 0.39 & 1.5 & 1.200 & 19.20 & 0.01 & 1.203 \\ 
B08 & N909 & 13.48 & 0.27 & 1.2 & 1.284 & 19.36 & 0.03 & 1.235 \\ 
B09 & N910 & 13.42 & 0.27 & 1.4 & 2.025 & 21.66 & 0.01 & 1.227 \\ 
B10 & N911 & 13.21 & 0.27 & 1.5 & 1.206 & 18.81 & 0.02 & 1.312 \\ 
B11 & N912 & 13.81 & 0.24 & 1.2 & 1.219 & 19.52 & 0.01 & 1.138 \\ 
B16 & CGCG539-042 & 13.88 & 0.30 & 1.6 & 1.336 & 19.94 & 0.03 & 1.120 \\& & & & & & & \\ 
{\bf Cluster : J8} \\ 
& & & & & & & \\ 
J01049 & CGCG483-070 & 13.93 & 0.31 & 1.5 & 1.164 & 19.32 & 0.01 & 1.136 \\ 
J01055 & - & 14.43 & 0.28 & 1.4 & 1.917 & 22.31 & 0.02 & 0.829 \\ 
& & 14.46 & 0.28 & 2.2 & 1.909 & 22.29 & 0.01 & 0.832 \\ 
& & 14.44 & 0.28 & 1.9 & 1.920 & 22.31 & 0.01 & 0.832 \\ 
J01056 & CGCG483-068 & 14.08 & 0.31 & 1.4 & 1.667 & 21.14 & 0.03 & 1.058 \\ 
& & 14.09 & 0.31 & 2.2 & 1.701 & 21.25 & 0.04 & 1.052 \\ 
& & 14.08 & 0.31 & 1.9 & 1.717 & 21.29 & 0.04 & 1.054 \\ 
J01060 & I1803 & 13.43 & 0.28 & 1.8 & 1.278 & 19.22 & 0.02 & 1.273 \\ 
J01065 & - & 14.83 & 0.28 & 2.0 & 0.856 & 19.01 & 0.01 & 0.912 \\ 
& & 14.82 & 0.28 & 1.0 & 0.829 & 18.90 & 0.02 & 0.914 \\ 
J01067 & EFAR-J8-I & 14.65 & 0.28 & 3.0 & 1.205 & 20.18 & 0.03 & 0.920 \\ 
J01069 & I1807 & 14.19 & 0.31 & 2.0 & 1.118 & 19.44 & 0.03 & 1.062 \\ 
J01070 & I1806 & 14.37 & 0.31 & 2.4 & 1.299 & 20.22 & 0.01 & 1.004 \\ 
& & 14.37 & 0.31 & 1.8 & 1.290 & 20.20 & 0.01 & 1.004 \\ 
J01080 & - & 15.09 & 0.34 & 1.8 & 0.869 & 19.28 & 0.01 & 0.855 \\ 
J03049 & - & 14.11 & 0.28 & 2.2 & 1.431 & 20.48 & 0.01 & 1.062 \\ 
& & 14.11 & 0.28 & 1.6 & 1.445 & 20.53 & 0.02 & 1.062 \\ 
J07038 & - & 15.07 & 0.32 & 2.4 & 0.991 & 19.74 & 0.01 & 0.846 \\ 
& & 15.06 & 0.32 & 1.0 & 0.995 & 19.74 & 0.01 & 0.850 \\ 
J08035 & - & 15.24 & 0.33 & 1.9 & 0.993 & 20.04 & 0.09 & 0.724 \\ 
& & 15.24 & 0.33 & 2.0 & 1.254 & 20.95 & 0.04 & 0.721 \\ 
J08036 & EFAR-J8-K & 14.53 & 0.33 & 1.9 & 1.215 & 20.06 & 0.02 & 0.972 \\ 
& & 14.53 & 0.33 & 2.0 & 1.197 & 19.99 & 0.02 & 0.972 \\ 
J09035 & - & 14.84 & 0.35 & 2.0 & 1.143 & 20.01 & 0.05 & 0.915 \\ 
& & & & & & & \\ 
{\bf Cluster : Perseus (A0426)} \\ 
& & & & & & & \\ 
P01 & I0293 & 13.93 & 0.56 & 0.9 & 1.597 & 20.76 & 0.03 & 1.114 \\ 
P02 & N1224 & 13.34 & 0.56 & 1.0 & 1.455 & 19.71 & 0.01 & 1.322 \\ 
P03 & I0310 & 13.20 & 0.60 & 1.0 & 1.628 & 20.06 & 0.01 & 1.377 \\ 
P05 & I0312 & 13.49 & 0.76 & 1.2 & 1.399 & 19.58 & 0.03 & 1.310 \\ 
P07 & CR19 & 14.69 & 0.65 & 1.3 & 1.336 & 20.61 & 0.02 & 0.912 \\ 
P08 & CR20 & 13.70 & 0.65 & 1.3 & 1.482 & 20.14 & 0.04 & 1.234 \\ 
P11 & BGP44 & 14.25 & 0.70 & 1.1 & 1.276 & 19.94 & 0.02 & 1.072 \\ 
P12 & N1270 & 13.00 & 0.65 & 1.2 & 1.164 & 18.23 & 0.02 & 1.419 \\ 
& & 13.01 & 0.65 & 1.3 & 1.138 & 18.15 & 0.01 & 1.416 \\ 
& & 13.01 & 0.65 & 1.5 & 1.145 & 18.17 & 0.02 & 1.416 \\ 
P13 & PER195 & 14.07 & 0.69 & 1.1 & 1.393 & 20.08 & 0.02 & 1.140 \\ 
& & 14.08 & 0.69 & 1.1 & 1.370 & 20.01 & 0.01 & 1.138 \\ 
P14 & PER199 & 14.10 & 0.69 & 1.1 & 1.052 & 18.88 & 0.02 & 1.147 \\ 
& & 14.10 & 0.69 & 1.1 & 1.012 & 18.73 & 0.01 & 1.146 \\ 
\end{tabular} 
\end{minipage} 
\end{table*} 

\begin{table*} 
\begin{minipage}{100mm} 
\scriptsize 
\contcaption{} 
\begin{tabular}{llrrrrrrr} 
& & & & & & & \\ 
Our ID & Other ID & $R_{20}$ & $A_{B}$ & psf & $\log A_{\rm{e}}$ & $\langle\mu\rangle_{\rm{e}}$ & rms & $\log D\sb{n}$ \\ 
P15 & CR28 & 14.12 & 0.70 & 1.2 & 1.097 & 19.04 & 0.03 & 1.140 \\ 
P16 & CR27 & 13.95 & 0.69 & 1.1 & 1.348 & 19.78 & 0.02 & 1.182 \\ 
& & 13.96 & 0.69 & 1.1 & 1.370 & 19.85 & 0.03 & 1.179 \\ 
P17 & N1272 & 13.09 & 0.65 & 1.2 & 1.775 & 20.39 & 0.03 & 1.414 \\ 
P18 & N1273 & 13.27 & 0.70 & 1.2 & 1.258 & 18.83 & 0.02 & 1.362 \\ 
P19 & I1907 & 13.66 & 0.70 & 1.1 & 1.460 & 20.01 & 0.04 & 1.246 \\ 
P20 & BGP111 & 15.41 & 0.69 & 1.3 & 0.764 & 19.06 & 0.03 & 0.802 \\ 
& & 15.41 & 0.69 & 1.2 & 0.900 & 19.59 & 0.02 & 0.799 \\ 
P21 & PER152 & 14.96 & 0.65 & 1.3 & 0.839 & 18.96 & 0.02 & 0.914 \\ 
& & 14.96 & 0.65 & 1.2 & 0.884 & 19.12 & 0.01 & 0.914 \\ 
P22 & CR36 & 14.04 & 0.70 & 1.1 & 1.095 & 18.93 & 0.02 & 1.169 \\ 
P23 & N1278 & 13.07 & 0.70 & 1.1 & 1.660 & 20.00 & 0.01 & 1.441 \\ 
& & 13.10 & 0.70 & 1.3 & 1.650 & 20.00 & 0.01 & 1.431 \\ 
& & 13.10 & 0.70 & 1.4 & 1.673 & 20.06 & 0.01 & 1.432 \\ 
P26 & BGP59 & 14.15 & 0.70 & 1.1 & 0.788 & 17.85 & 0.01 & 1.146 \\ 
& & 14.18 & 0.70 & 1.3 & 0.805 & 17.94 & 0.00 & 1.137 \\ 
& & 14.16 & 0.70 & 1.4 & 0.823 & 18.00 & 0.00 & 1.141 \\ 
P27 & U02673 & 13.68 & 0.69 & 0.9 & 1.559 & 20.27 & 0.04 & 1.235 \\ 
P28 & N1281 & 13.55 & 0.70 & 0.9 & 1.169 & 18.82 & 0.01 & 1.281 \\ 
P29 & N1282 & 13.20 & 0.69 & 1.4 & 1.406 & 19.34 & 0.02 & 1.368 \\ 
& & 13.19 & 0.69 & 1.1 & 1.409 & 19.34 & 0.02 & 1.371 \\ 
P30 & N1283 & 13.70 & 0.69 & 1.4 & 1.225 & 19.11 & 0.01 & 1.251 \\ 
& & 13.71 & 0.69 & 1.2 & 1.218 & 19.10 & 0.01 & 1.248 \\ 
P31 & PER163 & 14.55 & 0.77 & 1.9 & 0.828 & 18.37 & 0.00 & 1.057 \\ 
P33 & BGP33 & 14.03 & 0.77 & 1.2 & 1.157 & 19.16 & 0.02 & 1.168 \\ 
& & 14.00 & 0.77 & 1.4 & 1.151 & 19.11 & 0.02 & 1.176 \\ 
P34 & I0313 & 13.54 & 0.70 & 2.2 & 1.393 & 19.58 & 0.02 & 1.285 \\ 
P36 & N1293 & 13.45 & 0.84 & 2.1 & 1.279 & 19.03 & 0.01 & 1.331 \\ 
P37 & U02698 & 13.20 & 0.75 & 2.5 & 1.306 & 18.88 & 0.01 & 1.398 \\ 
P38 & U02717 & 13.55 & 0.85 & 2.5 & 1.436 & 19.68 & 0.01 & 1.306 \\ 
P39 & U02725 & 13.77 & 0.88 & 1.8 & 1.199 & 18.98 & 0.02 & 1.260 \\ 
& & & & & & & \\ 
{\bf Cluster : Coma (A1656)} \\ 
& & & & & & & \\ 
D68 & I3963 & 14.54 & 0.04 & 1.0 & 1.266 & 20.50 & 0.02 & 0.887 \\ 
D69 & I3959 & 14.04 & 0.04 & 1.0 & 1.074 & 19.27 & 0.01 & 1.065 \\ 
D70 & I3957 & 14.56 & 0.04 & 1.0 & 0.874 & 18.99 & 0.01 & 0.934 \\ 
D104 & N4875 & 14.39 & 0.04 & 1.3 & 0.803 & 18.51 & 0.01 & 0.998 \\ 
D105 & N4869 & 13.71 & 0.05 & 1.3 & 1.215 & 19.48 & 0.01 & 1.148 \\ 
D122 & N4894 & 14.78 & 0.05 & 1.2 & 0.993 & 19.72 & 0.01 & 0.852 \\ 
& & 14.80 & 0.05 & 1.2 & 0.936 & 19.51 & 0.01 & 0.852 \\ 
D124 & N4876 & 14.12 & 0.05 & 1.3 & 0.972 & 18.99 & 0.03 & 1.044 \\ 
D125 & - & 15.04 & 0.05 & 1.3 & 0.484 & 17.71 & 0.01 & 0.876 \\ 
D126 & - & 15.32 & 0.05 & 1.3 & 0.955 & 20.07 & 0.03 & 0.702 \\ 
D127 & - & 15.70 & 0.05 & 1.3 & 0.749 & 19.57 & 0.02 & 0.650 \\ 
D128 & - & 15.40 & 0.05 & 1.3 & 0.680 & 18.96 & 0.02 & 0.754 \\ 
D129 & N4874 & 13.14 & 0.05 & 1.3 & 2.112 & 21.68 & 0.04 & 1.303 \\ 
D130 & N4872 & 14.20 & 0.05 & 1.3 & 0.809 & 18.33 & 0.01 & 1.050 \\ 
D131 & N4871 & 14.16 & 0.05 & 1.3 & 1.144 & 19.66 & 0.02 & 1.012 \\ 
D132 & - & 15.30 & 0.05 & 1.3 & 0.905 & 19.85 & 0.01 & 0.727 \\ 
D148 & N4889 & 12.55 & 0.05 & 1.3 & 1.813 & 20.30 & 0.02 & 1.489 \\ 
& & 12.55 & 0.05 & 1.2 & 1.808 & 20.30 & 0.02 & 1.489 \\ 
D149 & - & 15.71 & 0.05 & 1.3 & 1.026 & 20.71 & 0.02 & 0.564 \\ 
& & 15.72 & 0.05 & 1.2 & 0.994 & 20.61 & 0.02 & 0.561 \\ 
D150 & I4011 & 14.76 & 0.05 & 1.3 & 0.983 & 19.65 & 0.02 & 0.865 \\ 
& & 14.77 & 0.05 & 1.2 & 0.992 & 19.69 & 0.02 & 0.861 \\ 
D151 & N4886 & 14.09 & 0.05 & 1.3 & 1.214 & 19.87 & 0.01 & 1.023 \\ 
& & 14.11 & 0.05 & 1.2 & 1.219 & 19.92 & 0.01 & 1.019 \\ 
D152 & I3998 & 14.52 & 0.05 & 1.3 & 1.110 & 19.89 & 0.03 & 0.914 \\ 
D154 & - & 15.55 & 0.05 & 1.3 & 1.319 & 21.72 & 0.01 & 0.488 \\ 
D155 & N4873 & 14.20 & 0.05 & 1.3 & 1.114 & 19.64 & 0.02 & 0.998 \\ 
D157 & - & 15.08 & 0.05 & 1.3 & 0.862 & 19.44 & 0.00 & 0.802 \\ 
& & & & & & & \\ 
{\bf Cluster : A2199} \\ 
& & & & & & & \\ 
A21-F113 & - & 15.45 & 0.00 & 1.3 & 0.617 & 18.74 & 0.00 & 0.750 \\ 
A21-F114 & - & 15.11 & 0.00 & 1.3 & 0.567 & 18.15 & 0.00 & 0.849 \\ 
A21-F121 & A2199-S26 & 14.43 & 0.00 & 1.7 & 1.265 & 20.38 & 0.00 & 0.924 \\ 
A21-F144 & A2199-S30 & 14.74 & 0.00 & 1.2 & 0.599 & 17.93 & 0.01 & 0.934 \\ 
& & 14.76 & 0.00 & 1.3 & 0.602 & 17.98 & 0.00 & 0.927 \\ 
A21-F145 & - & 14.56 & 0.00 & 1.2 & 1.235 & 20.42 & 0.03 & 0.874 \\ 
& & 14.60 & 0.00 & 1.3 & 1.182 & 20.28 & 0.02 & 0.866 \\ 
A21-F146 & A2199-S34 & 15.28 & 0.00 & 1.2 & 0.713 & 18.99 & 0.01 & 0.779 \\ 
& & 15.30 & 0.00 & 1.3 & 0.696 & 18.93 & 0.01 & 0.779 \\ 
A21-F164 & N6166 & 13.41 & 0.00 & 1.2 & 2.184 & 22.16 & 0.05 & 1.191 \\ 
& & 13.40 & 0.00 & 1.3 & 2.179 & 22.14 & 0.05 & 1.195 \\ 
& & 13.36 & 0.00 & 1.7 & 2.218 & 22.20 & 0.05 & 1.216 \\ 
A21-Z34A & A2199-Z34A & 14.08 & 0.00 & 1.3 & 1.251 & 19.95 & 0.03 & 1.035 \\ 
A21-Z34AC & - & 14.59 & 0.00 & 1.3 & 0.775 & 18.55 & 0.01 & 0.953 \\ 
\end{tabular} 
\end{minipage} 
\end{table*} 

\begin{table*} 
\begin{minipage}{100mm} 
\scriptsize 
\contcaption{} 
\begin{tabular}{llrrrrrrr} 
& & & & & & & \\ 
Our ID & Other ID & $R_{20}$ & $A_{B}$ & psf & $\log A_{\rm{e}}$ & $\langle\mu\rangle_{\rm{e}}$ & rms & $\log D\sb{n}$ \\ 
& & & & & & & \\ 
{\bf Cluster : A2634} \\ 
& & & & & & & \\ 
A26-F1201 & A2634-D79 & 15.18 & 0.18 & 1.6 & 0.791 & 19.11 & 0.01 & 0.823 \\ 
& & 15.17 & 0.18 & 2.5 & 0.803 & 19.16 & 0.01 & 0.819 \\ 
A26-F121 & A2634-D80 & 15.28 & 0.18 & 1.6 & 0.660 & 18.64 & 0.01 & 0.816 \\ 
& & 15.29 & 0.18 & 2.5 & 0.676 & 18.72 & 0.02 & 0.811 \\ 
A26-F1221 & N7720 & 13.34 & 0.16 & 1.2 & 1.592 & 20.26 & 0.04 & 1.272 \\ 
& & 13.34 & 0.16 & 1.6 & 1.568 & 20.17 & 0.05 & 1.273 \\ 
A26-F1222 & A2634-D76 & 14.62 & 0.16 & 1.3 & 0.788 & 18.59 & 0.02 & 0.963 \\ 
& & 14.62 & 0.16 & 1.6 & 0.751 & 18.44 & 0.02 & 0.964 \\ 
A26-F129 & A2634-D74 & 14.58 & 0.16 & 1.3 & 1.040 & 19.58 & 0.00 & 0.942 \\ 
& & 14.57 & 0.16 & 1.6 & 1.029 & 19.51 & 0.01 & 0.948 \\ 
A26-F134 & A2634-D55 & 14.25 & 0.16 & 1.2 & 1.112 & 19.51 & 0.02 & 1.031 \\ 
& & 14.24 & 0.16 & 1.6 & 1.096 & 19.43 & 0.02 & 1.036 \\ 
A26-F138 & A2634-D58 & 14.28 & 0.14 & 1.2 & 1.210 & 19.89 & 0.03 & 1.008 \\ 
A26-F139 & A2634-D57 & 14.13 & 0.14 & 1.2 & 1.218 & 19.81 & 0.01 & 1.050 \\ 
\end{tabular} 
\end{minipage} 
\end{table*} 

\clearpage 

\begin{table*} 
\begin{minipage}{170mm} 
\scriptsize 
\caption{ 
Combined spectroscopic and photometric parameters. For each galaxy 
with both spectroscopic and photometric data, we tabulate : Type = 
morphological type assigned from CCD images or other source (E = 
elliptical, S0 = S0/lenticular, R = morphological reject -- spiral, 
disky S0 etc. -- Q = unclassified); $cz$ = heliocentric recession 
velocity (kms$^{-1}$; $N_{\sigma}$ = number of velocity dispersion 
measurements; 
$\sigma$ = central velocity dispersion (kms$^{-1}$; corrected to standard 
system, see text); $\varepsilon_{\sigma}$ = poisson error on mean 
$\sigma$ value, $N\sb{Mg_{2}}$ = number of \mg\ measurements; \mg\ = 
magnesium index (magnitudes; corrected to standard system); 
$N_{\dn}$ = number of 
photometric observations; $A_{B}$ = B-band absorption coefficient; 
log$A_{\rm{e}}$ = log effective diameter (arcsec); 
$\langle\mu\rangle_{\rm{e}}$ = mean surface brightness within 
$A_{\rm{e}}$; log\dn\ = log R-band photometric \dn\ parameter (arcsec).} 

\label{mastable} 
\begin{tabular}{lllrcrrcrrcrrrr} 
& & & & & & & & & & & & & \\ 
Our ID & Other ID & Type & cz & $N_{\sigma}$ & $\log\sigma$ & $\varepsilon_{\log\sigma}$ & N$\sb{Mg_{2}}$ & \mg\ & $\varepsilon\sb{Mg_{2}}$ & $N_{D_{n}}$ & $A_{B}$ & $\log A_{\rm{e}}$ & $\langle\mu\rangle_{\rm{e}}$ & $\log\dn$ \\ 
& & & & & & & & & & & & & \\ 
\multicolumn{2}{l}{\bf Cluster : 7S21} \\ 
& & & & & & & & & & & & & \\ 
S01 & N0079 & E & 5479 & 1 & 2.280 & 0.030 & 1 & 0.312 & 0.009 & 1 & 0.05 & 1.354 & 20.00 & 1.132 \\ 
S02 & N0085A & S0 & 6189 & 1 & 2.025 & 0.030 & 1 & 0.244 & 0.009 & 1 & 0.05 & 1.502 & 20.99 & 0.953 \\ 
S03 & N0083 & E & 6262 & 2 & 2.395 & 0.021 & 2 & 0.326 & 0.009 & 1 & 0.09 & 1.659 & 20.43 & 1.318 \\ 
S04 & N0080 & E & 5741 & 2 & 2.398 & 0.021 & 2 & 0.308 & 0.006 & 1 & 0.09 & 1.691 & 20.28 & 1.364 \\ 
S05 & I1548 & S0 & 5775 & 1 & 2.165 & 0.030 & 1 & 0.202 & 0.009 & 1 & 0.09 & 0.992 & 18.91 & 1.072 \\ 
S06 & - & S0 & 5646 & 2 & 2.103 & 0.021 & 2 & 0.211 & 0.009 & 1 & 0.09 & 1.128 & 20.34 & 0.796 \\ 
S07 & CGCG457-008 & S0 & 5926 & 1 & 2.053 & 0.030 & 1 & 0.259 & 0.009 & 1 & 0.11 & 1.128 & 19.47 & 1.055 \\ 
& & & & & & & & & & & & & \\ 
\multicolumn{2}{l}{\bf Cluster : Pisces} \\ 
& & & & & & & & & & & & & \\ 
Z17005 & - & E & 4651 & 1 & 2.010 & 0.030 & 1 & 0.208 & 0.009 & 1 & 0.17 & 0.937 & 19.14 & 0.969 \\ 
Z16012 & - & R & 17740 & 1 & 2.224 & 0.030 & 1 & 0.293 & 0.009 & 1 & 0.14 & 0.941 & 19.80 & 0.779 \\ 
Z14028 & CGCG501-070 & E & 4263 & 2 & 2.294 & 0.024 & 2 & 0.327 & 0.007 & 2 & 0.18 & 0.928 & 18.40 & 1.143 \\ 
Z01034 & CGCG501-077 & E & 5156 & 2 & 2.079 & 0.024 & 2 & 0.270 & 0.007 & 1 & 0.16 & 1.192 & 19.79 & 1.026 \\ 
Z01036 & I1618 & S0 & 4730 & 1 & 1.948 & 0.040 & 1 & 0.226 & 0.012 & 2 & 0.16 & 1.247 & 20.19 & 0.965 \\ 
Z01047 & - & E & 5493 & 1 & 2.110 & 0.030 & 1 & 0.288 & 0.009 & 1 & 0.17 & 0.839 & 18.66 & 0.991 \\ 
Z01043 & N0375 & S0 & 5920 & 1 & 2.249 & 0.040 & 1 & 0.277 & 0.012 & 1 & 0.17 & 0.840 & 18.50 & 1.035 \\ 
Z01027 & N0379 & S0 & 5502 & 2 & 2.355 & 0.024 & 2 & 0.300 & 0.007 & 3 & 0.17 & 1.542 & 19.89 & 1.376 \\ 
Z01030 & N0380 & E & 4440 & 5 & 2.470 & 0.015 & 5 & 0.341 & 0.004 & 3 & 0.17 & 1.319 & 18.94 & 1.387 \\ 
Z01035 & N0383 & E & 5093 & 5 & 2.449 & 0.015 & 5 & 0.311 & 0.005 & 2 & 0.17 & 1.789 & 20.27 & 1.485 \\ 
Z01035C1 & N0382 & E & 5250 & 4 & 2.299 & 0.017 & 4 & 0.276 & 0.005 & 2 & 0.17 & 1.086 & 18.90 & 1.188 \\ 
Z01049 & N0384 & E & 4266 & 3 & 2.410 & 0.021 & 3 & 0.314 & 0.006 & 3 & 0.17 & 1.190 & 18.88 & 1.282 \\ 
Z01045 & N0385 & E & 5016 & 3 & 2.294 & 0.019 & 3 & 0.293 & 0.007 & 4 & 0.17 & 1.396 & 19.68 & 1.257 \\ 
Z01041 & N0386 & E & 5560 & 2 & 2.102 & 0.024 & 2 & 0.254 & 0.007 & 2 & 0.17 & 1.075 & 19.40 & 1.036 \\ 
Z01046 & N0388 & E & 5464 & 2 & 2.130 & 0.024 & 2 & 0.257 & 0.007 & 2 & 0.18 & 0.881 & 18.67 & 1.039 \\ 
Z01032 & - & S0 & 4763 & 1 & 2.011 & 0.040 & 1 & 0.274 & 0.012 & 1 & 0.18 & 1.100 & 20.08 & 0.852 \\ 
Z04050 & N0392 & E & 4679 & 3 & 2.361 & 0.021 & 3 & 0.302 & 0.006 & 3 & 0.18 & 1.374 & 19.29 & 1.351 \\ 
Z04049 & N0394 & S0 & 4388 & 3 & 2.257 & 0.021 & 3 & 0.270 & 0.006 & 3 & 0.18 & 1.052 & 18.76 & 1.184 \\ 
Z04051 & N0397 & E & 4988 & 1 & 2.083 & 0.030 & 1 & 0.261 & 0.009 & 3 & 0.18 & 1.016 & 19.28 & 1.000 \\ 
Z01026 & N0398 & S0 & 4912 & 1 & 2.006 & 0.030 & 1 & 0.264 & 0.009 & 1 & 0.18 & 1.155 & 19.63 & 1.046 \\ 
Z01073 & CGCG501-102 & E & 5176 & 2 & 2.187 & 0.024 & 2 & 0.280 & 0.007 & 2 & 0.18 & 1.061 & 19.17 & 1.062 \\ 
Z05052 & N0410 & E & 5314 & 4 & 2.470 & 0.016 & 4 & 0.352 & 0.006 & 2 & 0.19 & 1.735 & 19.93 & 1.524 \\ 
Z10020 & CGCG501-126 & S0 & 4852 & 1 & 1.919 & 0.030 & 1 & 0.230 & 0.009 & 1 & 0.20 & 1.134 & 20.08 & 0.879 \\ 
Z02057 & N0420 & E & 5026 & 2 & 2.258 & 0.024 & 2 & 0.249 & 0.007 & 4 & 0.16 & 1.531 & 19.88 & 1.336 \\ 
Z05034 & I1638 & S0 & 4828 & 4 & 2.153 & 0.016 & 4 & 0.278 & 0.005 & 2 & 0.16 & 1.244 & 19.63 & 1.119 \\ 
Z05044 & I1648 & S0 & 5541 & 1 & 2.083 & 0.030 & 1 & 0.269 & 0.009 & 1 & 0.17 & 1.116 & 19.47 & 1.050 \\ 
& & & & & & & & & & & & & \\ 
\multicolumn{2}{l}{\bf Cluster : HMS0122+3305} \\ 
& & & & & & & & & & & & & \\ 
H01056 & I1680 & S0 & 4438 & 1 & 2.122 & 0.030 & 1 & 0.269 & 0.009 & 1 & 0.17 & 1.035 & 19.05 & 1.073 \\ 
H01078 & I1673 & E & 5090 & 1 & 2.268 & 0.030 & 1 & 0.277 & 0.009 & 1 & 0.13 & 0.856 & 18.14 & 1.139 \\ 
H01051 & CGCG502-043 & E & 5237 & 2 & 2.097 & 0.024 & 1 & 0.268 & 0.009 & 1 & 0.13 & 1.100 & 19.41 & 1.048 \\ 
H01041 & N0499 & E & 4395 & 2 & 2.405 & 0.024 & 1 & 0.329 & 0.009 & 1 & 0.17 & 1.545 & 19.44 & 1.481 \\ 
H01044 & N0501 & E & 5010 & 1 & 2.201 & 0.030 & 1 & 0.306 & 0.009 & 1 & 0.16 & 1.029 & 19.16 & 1.041 \\ 
H01064 & N0507 & S0 & 4937 & 3 & 2.458 & 0.019 & 2 & 0.299 & 0.006 & 1 & 0.16 & 1.668 & 19.70 & 1.512 \\ 
H01057 & N0508 & E & 5517 & 2 & 2.343 & 0.024 & 1 & 0.312 & 0.009 & 1 & 0.16 & 1.445 & 19.96 & 1.224 \\ 
H01022 & N0528 & E & 4810 & 1 & 2.372 & 0.040 & 0 & - & - & 1 & 0.17 & 1.330 & 19.20 & 1.338 \\ 
H04010 & N0529 & E & 4796 & 2 & 2.363 & 0.024 & 1 & 0.297 & 0.009 & 1 & 0.15 & 1.424 & 19.14 & 1.448 \\ 
& & & & & & & & & & & & & \\ 
\multicolumn{2}{l}{\bf Cluster : A0262} \\ 
& & & & & & & & & & & & & \\ 
A01071 & N0705 & R & 4524 & 1 & 2.256 & 0.040 & 1 & 0.270 & 0.012 & 1 & 0.24 & 1.181 & 19.19 & 1.209 \\ 
A19041 & U01269 & S0 & 3848 & 2 & 2.055 & 0.021 & 2 & 0.187 & 0.006 & 1 & 0.13 & 1.479 & 21.05 & 0.910 \\ 
A01094 & - & S0 & 14620 & 1 & 2.433 & 0.030 & 1 & 0.312 & 0.009 & 1 & 0.18 & 1.030 & 19.48 & 0.955 \\ 
A01043 & N0687 & E & 5106 & 2 & 2.353 & 0.024 & 2 & 0.301 & 0.007 & 1 & 0.21 & 1.478 & 19.41 & 1.420 \\ 
A01076 & - & E & 4284 & 2 & 2.094 & 0.024 & 2 & 0.272 & 0.007 & 1 & 0.24 & 1.154 & 20.02 & 0.931 \\ 
A01074 & N0704 & S0 & 4724 & 2 & 2.194 & 0.024 & 2 & 0.288 & 0.007 & 1 & 0.24 & 1.153 & 19.22 & 1.155 \\ 
A01067 & N0703 & E & 5580 & 1 & 2.341 & 0.030 & 1 & 0.313 & 0.009 & 1 & 0.24 & 1.361 & 19.69 & 1.233 \\ 
A01069 & N0708 & E & 4864 & 2 & 2.340 & 0.021 & 2 & 0.321 & 0.006 & 1 & 0.24 & 2.095 & 21.75 & 1.282 \\ 
A01047 & CGCG522-048 & Q & 4151 & 1 & 2.147 & 0.030 & 1 & 0.265 & 0.009 & 1 & 0.24 & 1.327 & 19.95 & 1.104 \\ 
A09029 & I0171 & Q & 5381 & 2 & 2.280 & 0.024 & 2 & 0.258 & 0.007 & 1 & 0.20 & 1.777 & 20.47 & 1.407 \\ 
A02025 & N0759 & E & 4639 & 2 & 2.406 & 0.024 & 2 & 0.255 & 0.007 & 1 & 0.20 & 1.543 & 19.85 & 1.363 \\ 
A05096 & CGCG522-089 & E & 5245 & 1 & 1.953 & 0.030 & 1 & 0.222 & 0.009 & 1 & 0.18 & 1.290 & 20.55 & 0.897 \\ 
\end{tabular} 
\end{minipage} 
\end{table*} 

\begin{table*} 
\begin{minipage}{170mm} 
\scriptsize 
\contcaption{} 
\begin{tabular}{lllrcrrcrrcrrrr} 
& & & & & & & & & & & & & \\ 
\multicolumn{2}{l}{\bf Cluster : A0347} \\ 
& & & & & & & & & & & & & \\ 
B02 & U01837 & E & 6582 & 1 & 2.286 & 0.030 & 1 & 0.307 & 0.009 & 1 & 0.34 & 1.589 & 20.43 & 1.234 \\ 
B03 & U01841 & E & 6373 & 1 & 2.363 & 0.030 & 1 & 0.309 & 0.009 & 2 & 0.34 & 1.813 & 20.76 & 1.337 \\ 
B03C & - & Q & 6649 & 1 & 2.473 & 0.030 & 1 & 0.310 & 0.009 & 2 & 0.34 & 0.588 & 17.51 & 1.027 \\ 
B06 & U01859 & E & 5917 & 1 & 2.550 & 0.030 & 1 & 0.354 & 0.009 & 1 & 0.31 & 1.182 & 18.68 & 1.321 \\ 
B07 & CGCG538-065 & S0 & 5301 & 1 & 2.308 & 0.030 & 1 & 0.314 & 0.009 & 1 & 0.39 & 1.200 & 19.20 & 1.203 \\ 
B08 & N909 & E & 4978 & 1 & 2.273 & 0.030 & 1 & 0.276 & 0.009 & 1 & 0.27 & 1.284 & 19.36 & 1.235 \\ 
B09 & N910 & R & 5237 & 2 & 2.391 & 0.021 & 2 & 0.339 & 0.006 & 1 & 0.27 & 2.025 & 21.66 & 1.227 \\ 
B10 & N911 & S0 & 5766 & 1 & 2.400 & 0.030 & 1 & 0.328 & 0.009 & 1 & 0.27 & 1.206 & 18.81 & 1.312 \\ 
B11 & N912 & E & 4418 & 1 & 2.235 & 0.030 & 1 & 0.295 & 0.009 & 1 & 0.24 & 1.219 & 19.52 & 1.138 \\ 
B16 & CGCG539-042 & E & 4885 & 1 & 2.185 & 0.030 & 1 & 0.272 & 0.009 & 1 & 0.30 & 1.336 & 19.94 & 1.120 \\ 
& & & & & & & & & & & & & \\ 
\multicolumn{2}{l}{\bf Cluster : J8} \\ 
& & & & & & & & & & & & & \\ 
J07038 & - & S0 & 10136 & 1 & 2.261 & 0.030 & 1 & 0.284 & 0.009 & 2 & 0.32 & 0.993 & 19.74 & 0.848 \\ 
J09035 & - & S0 & 11133 & 1 & 2.451 & 0.030 & 1 & 0.340 & 0.009 & 1 & 0.35 & 1.143 & 20.01 & 0.915 \\ 
J08035 & - & R & 10099 & 1 & 1.910 & 0.030 & 1 & 0.182 & 0.009 & 2 & 0.33 & 1.124 & 20.49 & 0.722 \\ 
J08036 & EFAR-J8-K & E & 9803 & 3 & 2.288 & 0.021 & 3 & 0.297 & 0.006 & 2 & 0.33 & 1.206 & 20.02 & 0.972 \\ 
J01065 & - & S0 & 9103 & 1 & 2.125 & 0.030 & 1 & 0.168 & 0.009 & 2 & 0.28 & 0.843 & 18.95 & 0.913 \\ 
J01067 & EFAR-J8-I & E & 9237 & 2 & 2.300 & 0.024 & 2 & 0.315 & 0.007 & 1 & 0.28 & 1.205 & 20.18 & 0.920 \\ 
J01060 & I1803 & E & 9583 & 1 & 2.565 & 0.030 & 1 & 0.352 & 0.009 & 1 & 0.28 & 1.278 & 19.22 & 1.273 \\ 
J01070 & I1806 & E & 10219 & 3 & 2.332 & 0.021 & 3 & 0.316 & 0.006 & 2 & 0.31 & 1.295 & 20.21 & 1.004 \\ 
J03049 & - & E & 9933 & 2 & 2.400 & 0.024 & 2 & 0.275 & 0.007 & 2 & 0.28 & 1.438 & 20.51 & 1.062 \\ 
J01055 & - & E & 9616 & 2 & 2.184 & 0.021 & 2 & 0.294 & 0.006 & 3 & 0.28 & 1.915 & 22.30 & 0.831 \\ 
J01056 & CGCG483-068 & E & 9513 & 3 & 2.371 & 0.019 & 3 & 0.333 & 0.006 & 3 & 0.31 & 1.695 & 21.23 & 1.055 \\ 
J01049 & CGCG483-070 & E & 8562 & 3 & 2.494 & 0.021 & 3 & 0.322 & 0.006 & 1 & 0.31 & 1.164 & 19.32 & 1.136 \\ 
J01069 & I1807 & E & 9058 & 2 & 2.309 & 0.024 & 2 & 0.279 & 0.007 & 1 & 0.31 & 1.118 & 19.44 & 1.062 \\ 
J01080 & - & S0 & 9731 & 1 & 2.216 & 0.030 & 1 & 0.255 & 0.009 & 1 & 0.34 & 0.869 & 19.28 & 0.855 \\ 
& & & & & & & & & & & & & \\ 
\multicolumn{2}{l}{\bf Cluster : Perseus (A0426)} \\ 
& & & & & & & & & & & & & \\ 
P01 & I0293 & E & 4714 & 1 & 2.171 & 0.040 & 1 & 0.274 & 0.012 & 1 & 0.56 & 1.597 & 20.76 & 1.114 \\ 
P02 & N1224 & S0 & 5235 & 1 & 2.384 & 0.030 & 1 & 0.274 & 0.009 & 1 & 0.56 & 1.455 & 19.71 & 1.322 \\ 
P03 & I0310 & S0 & 5660 & 1 & 2.329 & 0.030 & 1 & 0.253 & 0.009 & 1 & 0.60 & 1.628 & 20.06 & 1.377 \\ 
P05 & I0312 & S0 & 4988 & 1 & 2.342 & 0.040 & 1 & 0.310 & 0.012 & 1 & 0.76 & 1.399 & 19.58 & 1.310 \\ 
P07 & CR19 & E & 3544 & 1 & 2.081 & 0.030 & 1 & 0.243 & 0.009 & 1 & 0.65 & 1.336 & 20.61 & 0.912 \\ 
P08 & CR20 & E & 6461 & 2 & 2.294 & 0.021 & 2 & 0.269 & 0.006 & 1 & 0.65 & 1.482 & 20.14 & 1.234 \\ 
P11 & BGP44 & E & 4247 & 1 & 2.192 & 0.030 & 1 & 0.279 & 0.009 & 1 & 0.70 & 1.276 & 19.94 & 1.072 \\ 
P12 & N1270 & E & 4997 & 2 & 2.530 & 0.024 & 2 & 0.361 & 0.007 & 3 & 0.65 & 1.149 & 18.18 & 1.417 \\ 
P13 & PER195 & E & 8396 & 2 & 2.231 & 0.024 & 2 & 0.286 & 0.007 & 2 & 0.69 & 1.382 & 20.05 & 1.139 \\ 
P14 & PER199 & S0 & 5102 & 3 & 2.323 & 0.019 & 3 & 0.289 & 0.006 & 2 & 0.69 & 1.032 & 18.81 & 1.147 \\ 
P15 & CR28 & E & 6213 & 1 & 2.317 & 0.030 & 1 & 0.292 & 0.009 & 1 & 0.70 & 1.097 & 19.04 & 1.140 \\ 
P16 & CR27 & S0 & 8063 & 1 & 2.228 & 0.040 & 1 & 0.280 & 0.012 & 2 & 0.69 & 1.359 & 19.82 & 1.180 \\ 
P17 & N1272 & S0 & 3801 & 7 & 2.417 & 0.013 & 6 & 0.346 & 0.004 & 1 & 0.65 & 1.775 & 20.39 & 1.414 \\ 
P18 & N1273 & S0 & 5397 & 1 & 2.311 & 0.040 & 1 & 0.263 & 0.012 & 1 & 0.70 & 1.258 & 18.83 & 1.362 \\ 
P19 & I1907 & S0 & 4489 & 1 & 2.285 & 0.040 & 1 & 0.292 & 0.012 & 1 & 0.70 & 1.460 & 20.01 & 1.246 \\ 
P20 & BGP111 & E & 3963 & 1 & 1.925 & 0.030 & 1 & 0.283 & 0.009 & 2 & 0.69 & 0.832 & 19.33 & 0.801 \\ 
P21 & PER152 & E & 3940 & 2 & 2.134 & 0.021 & 2 & 0.316 & 0.006 & 2 & 0.65 & 0.861 & 19.04 & 0.914 \\ 
P22 & CR36 & E & 7470 & 1 & 2.301 & 0.040 & 1 & 0.294 & 0.012 & 1 & 0.70 & 1.095 & 18.93 & 1.169 \\ 
P23 & N1278 & E & 6064 & 3 & 2.410 & 0.021 & 2 & 0.309 & 0.007 & 3 & 0.70 & 1.661 & 20.02 & 1.435 \\ 
P26 & BGP59 & E & 5315 & 1 & 2.307 & 0.030 & 1 & 0.287 & 0.009 & 3 & 0.70 & 0.805 & 17.93 & 1.141 \\ 
P27 & U02673 & E & 4434 & 1 & 2.290 & 0.040 & 1 & 0.302 & 0.012 & 1 & 0.69 & 1.559 & 20.27 & 1.235 \\ 
P28 & N1281 & E & 4300 & 1 & 2.432 & 0.030 & 1 & 0.328 & 0.009 & 1 & 0.70 & 1.169 & 18.82 & 1.281 \\ 
P29 & N1282 & E & 2223 & 4 & 2.325 & 0.017 & 4 & 0.290 & 0.005 & 2 & 0.69 & 1.408 & 19.34 & 1.370 \\ 
P30 & N1283 & E & 6746 & 2 & 2.320 & 0.028 & 1 & 0.291 & 0.012 & 2 & 0.69 & 1.222 & 19.11 & 1.249 \\ 
P31 & PER163 & E & 5481 & 2 & 2.249 & 0.021 & 2 & 0.290 & 0.006 & 1 & 0.77 & 0.828 & 18.37 & 1.057 \\ 
P33 & BGP33 & S0 & 4950 & 1 & 2.216 & 0.030 & 1 & 0.293 & 0.009 & 2 & 0.77 & 1.154 & 19.13 & 1.172 \\ 
P34 & I0313 & S0 & 4432 & 1 & 2.375 & 0.030 & 1 & 0.335 & 0.009 & 1 & 0.70 & 1.393 & 19.58 & 1.285 \\ 
P36 & N1293 & E & 4167 & 3 & 2.321 & 0.021 & 3 & 0.318 & 0.006 & 1 & 0.84 & 1.279 & 19.03 & 1.331 \\ 
P37 & U02698 & E & 6451 & 2 & 2.557 & 0.024 & 2 & 0.340 & 0.007 & 1 & 0.75 & 1.306 & 18.88 & 1.398 \\ 
P38 & U02717 & E & 3793 & 2 & 2.189 & 0.024 & 2 & 0.239 & 0.007 & 1 & 0.85 & 1.436 & 19.68 & 1.306 \\ 
P39 & U02725 & S0 & 6215 & 1 & 2.333 & 0.030 & 1 & 0.297 & 0.009 & 1 & 0.88 & 1.199 & 18.98 & 1.260 \\ 
& & & & & & & & & & & & & \\ 
\multicolumn{2}{l}{\bf Cluster : A2199} \\ 
& & & & & & & & & & & & & \\ 
A21-F113 & - & Q & 8052 & 4 & 2.217 & 0.015 & 4 & 0.269 & 0.004 & 1 & 0.00 & 0.617 & 18.74 & 0.750 \\ 
A21-F114 & - & S0 & 9184 & 4 & 2.297 & 0.015 & 4 & 0.303 & 0.004 & 1 & 0.00 & 0.567 & 18.15 & 0.849 \\ 
A21-F121 & A2199-S26 & E & 8768 & 2 & 2.239 & 0.021 & 2 & 0.287 & 0.006 & 1 & 0.00 & 1.265 & 20.38 & 0.924 \\ 
A21-F144 & A2199-S30 & E & 8524 & 2 & 2.411 & 0.021 & 2 & 0.262 & 0.006 & 2 & 0.00 & 0.601 & 17.95 & 0.931 \\ 
A21-F145 & - & R & 7610 & 2 & 2.173 & 0.021 & 2 & 0.289 & 0.006 & 2 & 0.00 & 1.209 & 20.35 & 0.870 \\ 
A21-F146 & A2199-S34 & E & 8314 & 2 & 2.205 & 0.021 & 2 & 0.272 & 0.006 & 2 & 0.00 & 0.704 & 18.96 & 0.779 \\ 
A21-F164 & N6166 & E & 9348 & 2 & 2.442 & 0.021 & 2 & 0.323 & 0.006 & 3 & 0.00 & 2.194 & 22.17 & 1.201 \\ 
A21-Z34A & A2199-Z34A & E & 8721 & 2 & 2.298 & 0.021 & 2 & 0.288 & 0.006 & 1 & 0.00 & 1.251 & 19.95 & 1.035 \\ 
A21-Z34AC & - & S0 & 8964 & 2 & 2.337 & 0.021 & 2 & 0.324 & 0.006 & 1 & 0.00 & 0.775 & 18.55 & 0.953 \\ 
\end{tabular} 
\end{minipage} 
\end{table*} 

\begin{table*} 
\begin{minipage}{170mm} 
\scriptsize 
\contcaption{} 
\begin{tabular}{lllrcrrcrrcrrrr} 
& & & & & & & & & & & & & \\ 
\multicolumn{2}{l}{\bf Cluster : A2634} \\ 
& & & & & & & & & & & & & \\ 
A26-F138 & A2634-D58 & E & 10872 & 4 & 2.343 & 0.017 & 4 & 0.316 & 0.005 & 1 & 0.14 & 1.210 & 19.89 & 1.008 \\ 
A26-F139 & A2634-D57 & E & 9591 & 4 & 2.331 & 0.017 & 4 & 0.327 & 0.005 & 1 & 0.14 & 1.218 & 19.81 & 1.050 \\ 
A26-F129 & A2634-D74 & S0 & 8425 & 3 & 2.315 & 0.019 & 3 & 0.301 & 0.006 & 2 & 0.16 & 1.035 & 19.55 & 0.945 \\ 
A26-F121 & A2634-D80 & E & 9582 & 4 & 2.262 & 0.017 & 4 & 0.286 & 0.005 & 2 & 0.18 & 0.668 & 18.68 & 0.814 \\ 
A26-F1201 & A2634-D79 & S0 & 10166 & 4 & 2.234 & 0.019 & 4 & 0.295 & 0.005 & 2 & 0.18 & 0.797 & 19.13 & 0.821 \\ 
A26-F134 & A2634-D55 & E & 9288 & 3 & 2.342 & 0.019 & 3 & 0.300 & 0.006 & 2 & 0.16 & 1.104 & 19.47 & 1.034 \\ 
A26-F129 & A2634-D74 & S0 & 8425 & 3 & 2.315 & 0.019 & 3 & 0.301 & 0.006 & 2 & 0.16 & 1.035 & 19.55 & 0.945 \\ 
A26-F1221 & N7720 & E & 9104 & 5 & 2.527 & 0.015 & 5 & 0.331 & 0.004 & 2 & 0.16 & 1.580 & 20.22 & 1.273 \\ 
A26-F1222 & A2634-D76 & E & 8123 & 5 & 2.320 & 0.015 & 5 & 0.296 & 0.004 & 2 & 0.16 & 0.770 & 18.52 & 0.964 \\ 
A26-F139 & A2634-D57 & E & 9591 & 4 & 2.331 & 0.017 & 4 & 0.327 & 0.005 & 1 & 0.14 & 1.218 & 19.81 & 1.050 \\

\end{tabular} 
\end{minipage} 
\end{table*}


\begin{thebibliography}{} 

\bibitem[\protect\citename{Andreon }1994]{and94} 
Andreon S. 1994 A\& A, 284, 801 

\bibitem[\protect\citename{Bower et al. }1992]{ble} 
Bower R.G., Lucey J.R., Ellis R.S. 1992, MNRAS, 254, 589 

\bibitem[\protect\citename{Bucknell, Godwin \& Peach }1979]{bgp} 
Bucknell M.J., Godwin J.G., Peach J.V. 1979, MNRAS, 188, 579 

\bibitem[\protect\citename{Burstein \& Heiles }1984]{bh} 
Burstein D., Heiles C. 1984, ApJS, 54, 33 

\bibitem[\protect\citename{Burstein et al. }1987]{bur87} 
Burstein D., Davies R.L., Dressler A., Faber S.M., Stone R.P.S., 
Lynden-Bell D., Terlevich R.J., Wegner G. 1987, ApJS, 64, 601 

\bibitem[\protect\citename{Chincarini \& Rood }1971]{cr71} 
Chincarini G., Rood H.J. 1971, ApJ, 168, 321 

\bibitem[\protect\citename{Colless et al. }1993]{col93} 
Colless M., Burstein D., Wegner G., Saglia R.P., McMahon R.K., 
Davies R.L., Bertschinger E., Baggley G. 1993, MNRAS, 262, 475 

\bibitem[\protect\citename{Davies et al. }1987]{dav87} 
Davies R.L., Burstein D., Dressler A., Faber S.M., Lynden-Bell D., 
Terlevich R.J., Wegner G. 1987, ApJS, 64, 581 

\bibitem[\protect\citename{Dekel }1994]{dek94} 
Dekel A. 1994, ARA\& A, 32, 371 

\bibitem[\protect\citename{Dressler }1980]{d80} 
Dressler A. 1980, ApJ, 236, 351 

\bibitem[\protect\citename{Dressler }1984]{d84} 
Dressler A. 1984, ApJ, 281, 512 

\bibitem[\protect\citename{Dressler et al. }1987]{dre87} 
Dressler A., Lynden-Bell D., Burstein D., Davies R.L., Burstein D., 
Faber S.M., Terlevich R.J., Wegner G. 1987, ApJ, 313, 42 

\bibitem[\protect\citename{Dressler et al. }1991]{dfb91} 
Dressler A., Faber S.M., Burstein D. 1991, ApJ, 368, 54 

\bibitem[\protect\citename{Faber et al. }1989]{fab89} 
Faber S.M., Wegner G., Burstein D., Davies R.L., Dressler A., 
Lynden-Bell D., Terlevich R.J. 1989, ApJS, 69, 763 

\bibitem[\protect\citename{Frei \& Gunn }1994]{frei94} 
Frei Z., Gunn J.E. AJ, 108, 1476 

\bibitem[\protect\citename{Gonzalez }1993]{gonza} 
Gonz\'alez J.J. 1993, PhD thesis, University of California, Santa 
Cruz 

\bibitem[\protect\citename{Guzman et al. }1992]{guz92} 
Guzm\'an R., Lucey J.R., Carter D., Terlevich R.J. 1992, MNRAS, 257, 
187 

\bibitem[\protect\citename{Gregg }1995]{gregg95} 
Gregg M.D. 1995, ApJ, 443, 527 

\bibitem[\protect\citename{Han \& Mould }1992]{han} 
Han M.-S., Mould J.R. 1992, ApJ, 396, 453 

\bibitem[\protect\citename{Huchra }1993]{zcat} 
Huchra J.P., Geller M.J., Clemens C.M., Tokarz S.P., Michel A. 1993, 
Astronomical Data Center archives 

\bibitem[\protect\citename{Hudson }1993]{hud93} 
Hudson M.J., 1993, MNRAS, 265, 43 

\bibitem[\protect\citename{Hudson }1994]{hud94} 
Hudson M.J., 1994, MNRAS, 266, 468 

\bibitem[\protect\citename{Hudson }1997]{pp2} 
Hudson M.J., Lucey J.R., Smith R.J., Steel J. 1997, MNRAS, submitted 

\bibitem[\protect\citename{Irwin \& McMahon}1992]{irwin92} 
Irwin M., McMahon R. 1992, Gemini (Newsletter of the Royal Greenwich Observatory), 37, 1 

\bibitem[\protect\citename{Jackson}1982]{jackson} 
Jackson R. 1982, Ph.D. thesis, University of California, Santa Cruz 

\bibitem[\protect\citename{J\o rgensen et al. }1995a]{jfkphot} 
J\o rgensen I, Franx M., Kj\ae rgaard P. 1995a MNRAS, 273, 1097 

\bibitem[\protect\citename{J\o rgensen et al. }1995b]{jfkspec} 
J\o rgensen I, Franx M., Kj\ae rgaard P. 1995b MNRAS, 276, 1341 

\bibitem[\protect\citename{J\o rgensen et al. }1996]{jfkfp} 
J\o rgensen I, Franx M., Kj\ae rgaard P. 1996 MNRAS, 280, 167 

\bibitem[\protect\citename{Kolatt \& Dekel }1994]{koldek} 
Kolatt T., Dekel A. 1994, ApJ, 428, 35 

\bibitem[\protect\citename{Landolt }1983]{land83} 
Landolt A.U. 1983, AJ, 88, 439 

\bibitem[\protect\citename{Landolt }1992]{land92} 
Landolt A.U. 1992, AJ, 104, 340. 

\bibitem[\protect\citename{Lucey \& Carter }1988]{lc} 
Lucey J.R., Carter D. 1988, MNRAS, 235, 1177 

\bibitem[\protect\citename{Lucey et al. }1991a]{lgyct} 
Lucey J.R., Gray P.M., Carter D., Terlevich R.J. 1991a, MNRAS, 248, 
804 

\bibitem[\protect\citename{Lucey et al. }1991b]{lgzct} 
Lucey J.R., Guzm\'an R., Carter D., Terlevich R.J. 1991b, MNRAS, 
253, 584 

\bibitem[\protect\citename{Lucey et al. }1993]{lgsc93} 
Lucey J.R., Guzm\'an R., Steel J., Carter D. 1994, in Cosmic 
Velocity Fields, ed. F. Bouchet \& M. Lachi\'eze-Rey, 
(Gif-sur-Yvette: Editions Fronti\`eres), p. 43 

\bibitem[\protect\citename{Lucey et al. }1997]{lgsc97} 
Lucey J.R., Guzm\'an R., Steel J., Carter D. 1997, MNRAS, 287, 899 

\bibitem[\protect\citename{Lucey et al. }1998]{focp2} 
Lucey J.R., Lahav O., Lynden-Bell D., Terlevich R.J., Infante, L., Melnick J. 
1998, in preparation 

\bibitem[\protect\citename{Lynden-Bell et al. }1988]{lb88} 
Lynden-Bell D., Faber S.M., Burstein D., Davies R.L., Dressler A., 
Terlevich R.J., Wegner G. 1988, ApJ, 326, 19 


\bibitem[\protect\citename{Oke \& Sandage}1968]{oke68} 
Oke J.B., Sandage A. 1968, ApJ., 154, 21 

\bibitem[\protect\citename{Postman \& Lauer }1995]{pl} 
Postman M., Lauer T.R. 1995, ApJ, 440, 28 (PL) 

\bibitem[\protect\citename{Poulain et al. }1992]{pnd92} 
Poulain P., Nieto J.-L., Davoust E. 1991, A\& AS, 95, 129 

\bibitem[\protect\citename{Saglia et al. }1997]{efar4} 
Saglia R.P., Bertschinger E., Baggley G., Burstein D., Colless M.M., 
Davies R.L., McMahan R.K., Wegner G. 1997, ApJS, 109, 79 

\bibitem[\protect\citename{Sargent et al. }1977]{sar77} 
Sargent W.L.W., Schechter P.L., Boksenberk A., Shortridge K. 1977, 
ApJ, 212, 326 

\bibitem[\protect\citename{Saunders et al. }1991]{sau91} 
Saunders W. et al. 1991, Nature, 349, 32 

\bibitem[\protect\citename{Steel }1997]{steel} 
Steel, J. 1997, Ph.D. thesis, University of Durham 

\bibitem[\protect\citename{Strauss \& Willick }1995]{sw} 
Strauss M.A., Willick J.A. 1995 Physics Reports, 261, 271 

\bibitem[\protect\citename{Tully \& Fisher }1977]{tf} 
Tully R.B., Fisher J.R. 1977, ApJ, A\& A, 54, 661 

\bibitem[\protect\citename{Wegner et al. }1996]{efar1} 
Wegner G., Colless M., Baggley G., Davies R.L., Bertschinger E., 
Burstein D., McMahan R.K., Saglia R.P. 1996, ApJS, 106, 1 

\bibitem[\protect\citename{Willick }1990]{w90} 
Willick J.A. 1990, ApJ, 352, L45 

\bibitem[\protect\citename{Willick }1991]{w91} 
Willick J.A. 1991, PhD thesis, University of California, Berkeley 

\bibitem[\protect\citename{Willick et al. }1997]{mk3} 
Willick J.A., Courteau S., Faber S.M., Burstein D., Dekel A., 
Strauss M.A. 1997, ApJS, 109, 333 


\end{thebibliography}
\end{document}